%% file: main.tex
\documentclass[sigconf,screen,authorversion,nonacm]{acmart}
\pdfoutput=1

\usepackage[utf8]{inputenc}
\usepackage{tcolorbox}
\usepackage{graphicx}
\usepackage{multirow,makecell}
\usepackage[ruled,vlined,linesnumbered]{algorithm2e}
\usepackage{subcaption}
\usepackage{booktabs}
\usepackage{pifont}
\usepackage{lipsum}
\usepackage{makecell}
\usepackage[para,online,flushleft]{threeparttable}
\usepackage{mathtools}
\usepackage{titlesec}
\usepackage{listings}
\usepackage{wrapfig}
\usepackage{balance}
\usepackage{amsmath}
\usepackage{adjustbox}
\usepackage{tabularx}
\usepackage[capitalise,noabbrev]{cleveref}
\usepackage{fancybox}
\usepackage[framemethod=tikz]{mdframed}
\usepackage{enumitem}
\usepackage{scalerel}

\usetikzlibrary{shadows}
\tikzset{every shadow/.style={opacity=1}}
\newmdenv[shadow=false,shadowcolor=black,shadowsize=0pt,linewidth=1pt,skipabove=2pt]{highlightbox}

\usepackage{hyperref}
\hypersetup{
    colorlinks=true,
    linkcolor=blue,
    filecolor=magenta,      
    urlcolor=cyan,
}

\titlespacing*{\section}{0pt}{10pt minus 2pt}{3pt}
\titlespacing*{\subsection}{0pt}{5pt minus 2pt}{3pt}

\title{Revisiting Query Performance in GPU Database Systems}
\author{Jiashen Cao$^{2,*}$, Rathijit Sen$^{1}$, Matteo Interlandi$^{1}$, Joy Arulraj$^{2}$, Hyesoon Kim$^{2}$}
\affiliation{
    \institution{$^{1}$Microsoft Gray Systems Lab, $^{2}$Georgia Institute of Technology}
    \country{USA}
}

\input{cmd}

\begin{document}

\input{abstract}

\settopmatter{printacmref=false, printccs=false, printfolios=false}

\maketitle
{
\let\thefootnote\relax\footnotetext{$^{*}$Work done during an internship at Microsoft Gray Systems Lab.}
}

\setcounter{page}{1}

\input{introduction}
\input{background}
\input{setup}
\input{perfanalysis}
\input{perfmodel}
\input{concurrency}
\input{evaluation}
\input{related}
\input{discuss}
\input{conclusion}

\clearpage
\bibliographystyle{ACM-Reference-Format}
\bibliography{ref}

\end{document}

%% file: cmd.tex
\SetCommentSty{mycommfont}

\newcommand{\probP}{\text{I\kern-0.15em P}}

\titlespacing*{\section}{0pt}{10pt minus 2pt}{3pt}
\titlespacing*{\subsection}{0pt}{5pt minus 2pt}{3pt}

\def\Snospace~{\S{}}

\newcommand{\crystal}{\textsc{Crystal}\xspace}
\newcommand{\heavydb}{\textsc{HeavyDB}\xspace}
\newcommand{\blzsql}{\textsc{BlazingSQL}\xspace}
\newcommand{\surakav}{\textsc{TQP}\xspace}
\newcommand{\pgstrom}{\textsc{PG-Strom}\xspace}

\newcommand{\eat}[1]{}

\newcommand{\eg}{\textit{e.g.,}\xspace}
\newcommand{\ie}{\textit{i.e.,}\xspace}
\newcommand{\vs}{\textit{vs.}\xspace}
\newcommand{\etal}{\textit{et al.}\xspace}

\newcommand{\PP}[1]{\vspace{4px}\noindent{\bf\textsc{#1.}}\xspace}

\def\Snospace~{\S{}}

\newcommand\BeraMonottfamily{%
  \def\fvm@Scale{0.85}
  \fontfamily{fvm}\selectfont
}

\lstdefinestyle{SQLStyle}{
  language=SQL,
  basicstyle={\small\ttfamily},
  breaklines=true,
  float=tp,
  floatplacement=tbp,  
  numbers=none,
  keepspaces=true,
  captionpos=b,
  aboveskip=50pt,
  belowskip=20pt,
  numberstyle=\tiny\color{gray},  
  stringstyle=\color{mauve},
  keywordstyle=\color{blue},
  commentstyle=\color{dkgreen},
  keywords=[2]{FasterRCNNResnet50, CarRecognitionResnet50, ColorDetector, LogicalObjectDetector},
  keywordstyle=[2]\color{bubbles},
}

\lstdefinestyle{SQLStyle}{
  language=SQL,
  basicstyle=\ttfamily\small, 
  keywordstyle=\color{codepurple}\bfseries,
  showstringspaces=false,
  aboveskip = 0.1in,
  belowskip = 0.05in,
  keywordstyle=\color{blue},  
  literate = {-}{-}1, 
}

\makeatletter
\newcommand*{\rom}[1]{\expandafter\@slowromancap\romannumeral #1@}
\makeatother

\newcommand{\squishitemize}{
 \begin{list}{$\bullet$}
  { \setlength{\itemsep}{0pt}
     \setlength{\parsep}{3pt}
     \setlength{\topsep}{3pt}
     \setlength{\partopsep}{0pt}
     \setlength{\leftmargin}{1.95em}
     \setlength{\labelwidth}{1.5em}
     \setlength{\labelsep}{0.5em} } }

\newcounter{Lcount}
\newcommand{\squishlist}{
    \begin{list}{\arabic{Lcount}. }
   { \usecounter{Lcount}
        \setlength{\itemsep}{0pt}
        \setlength{\parsep}{3pt}
        \setlength{\topsep}{3pt}
        \setlength{\partopsep}{0pt}
        \setlength{\leftmargin}{2em}
        \setlength{\labelwidth}{1.5em}
        \setlength{\labelsep}{0.5em} } }

\newcommand{\squishend}{\end{list}}

%% file: abstract.tex
\begin{abstract}

GPUs offer massive compute parallelism and high-bandwidth memory accesses.
GPU database systems seek to exploit those capabilities to accelerate data analytics. 
Although modern GPUs have more resources (\eg higher DRAM bandwidth) than ever before, 
judicious choices for query processing that avoid wasteful resource allocations are still advantageous.
Database systems can save 
GPU runtime costs through just-enough resource allocation or improve query 
throughput with
concurrent query processing by leveraging new GPU capabilities, such as Multi-Instance GPU (MIG).

In this paper we do a cross-stack performance and resource utilization analysis of five GPU database systems.
We study both database-level and micro-architectural aspects, and offer recommendations to database developers. 
We also demonstrate how to use and extend the traditional roofline model to identify GPU resource bottlenecks.
This enables users to conduct what-if analysis to forecast performance impact for different resource allocation or the degree of concurrency.
Our methodology addresses a key user pain point in selecting optimal configurations 
by removing the need to do exhaustive testing for a multitude of resource configurations.

\end{abstract}

%% file: introduction.tex
\section{INTRODUCTION}~\label{sec:intro}
Graphics Processing Units (GPUs), with their potential for massively parallel computing, high-bandwidth memory access capability, and relative easy of programming for an accelerator, have seen rising interests in their usage for accelerating data analytics, with a number of GPU database systems being developed in both academic and industrial settings in recent years~\cite{surakav_he_2022,blazingsql_2022,heavydb_2022,crystal_shanbhag_2020,pgstrom_2022,virginian_peter_2010,aresdb_shen_2019,r3d3_krolik_2021,gpl_paul_2016,pipeline_funke_2018,hwoblivious_heimel_2013,redfox_wu_2014,nestgpu_floratos_2021}. 
Recent advances in interconnect protocols~\cite{pcie4,pcie5,pcie6,pcie7} and architectural enhancements have made GPUs more attractive as accelerators for data analytics, and we expect an increase in the popularity of GPU database systems and a proliferation of research into improving their efficiency going forward.

Query performance in GPU database systems depends on several factors:
(1) the computational capacity of the GPU,
(2) the implementation of the query execution engine,
(3) the characteristics of the query, and 
(4) the database size.
A deeper understanding of GPU resource utilization and bottlenecks encountered is important for designing better GPU database systems. 

\PP{Prior Work}
While previous studies (e.g.,~\cite{gpuchar_furst_2017,gpuchar_suh_2022,surakav_he_2022}) have compared query performance across GPU database systems, a cross-stack analysis that connects query processing design choices with the impact on micro-architectural performance metrics is missing---conversely similar studies~\cite{cpudbanalysis_anastassia_1999,cpudbanalysis_sirin_2020,sqlresourcesensitivity_sen_2018} exist for CPU database systems. 
We aim to address this gap in this paper, 
and develop recommendations for GPU database developers and hardware vendors.

\begin{figure}[t]
\begin{subfigure}[t]{0.49\columnwidth}
 \centering\hspace{-5ex}
 \includegraphics[width=\columnwidth]{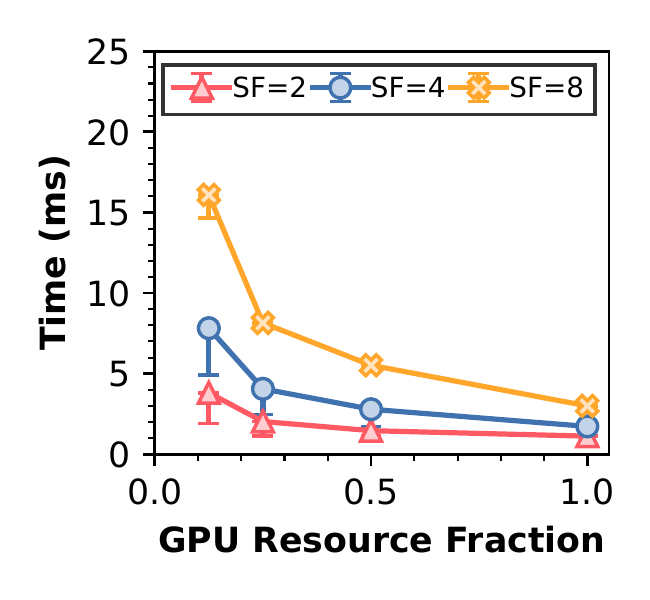}\vspace{-3ex}
 \caption{Same query (Q33) with different SFs.}
 \label{subfig:resource-scaling-sf}
\end{subfigure}
\hfill
\begin{subfigure}[t]{0.49\columnwidth}
 \centering
 \includegraphics[width=\columnwidth]{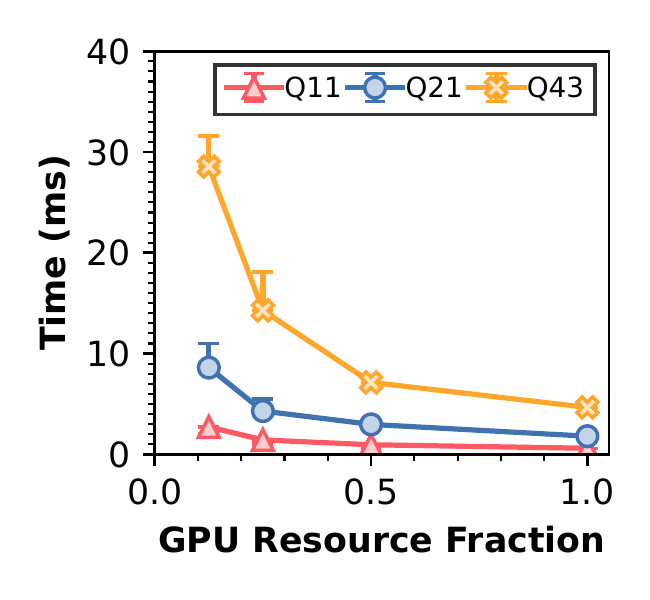}\vspace{-3ex}
 \caption{Different queries with same SF=4.}
 \label{subfig:resource-scaling-queries}
\end{subfigure}
\vspace{1ex}
\caption{
\textbf{Query Execution Time (and Estimation Error) \vs Resource Allocation --}
Query execution time \vs GPU resource allocations for representative queries (\autoref{sec:exp:query}) and scale factors (SFs), along with the error of the time estimates (shown with bars) using our predictive framework. X-axis denotes the fraction of GPU resources being used to run the queries.
}\vspace{-5pt}
\label{fig:resource-scaling-time}
\end{figure}

\PP{Cloud GPUs}
Several GPU instance types are now available on the cloud: from GPUs well fitting ML inference workloads (e.g., NVIDIA T4), to more powerful GPUs for ML training (e.g., P100, V100), to high-end one such as the A100.
Each of such GPUs have different performance/cost trade offs which are widely unexplored for database workloads.
For example, 
a T4 GPU can run the TPC-H benchmark 20\% slower compared to a P100, but at $\frac{1}{5}$ of the cost~\cite{surakav_he_2022}.
To complicate things further, starting from the Ampere generation, NVIDIA now allows 
different resource allocations strategies through the Multi-Instance GPU (MIG) capability~\cite{mig_capability_nvidia_2022}. This new capability allows to partition GPU hardware resources 
to either increase concurrency, or decrease the cost by sharing resources.
For instance, \cref{fig:resource-scaling-time} shows the query execution time and GPU
resource 
trade-offs for running representative queries from the Star Schema Benchmark (SSB)~\cite{ssb_rabl_2013} 
using \heavydb (an open-source GPU database system~\cite{heavydb_2022})
on an NVIDIA A100 GPU with MIG. 
We observe that query performance does not always scale linearly with respect to the allocated GPU resources. For example, performance of memory-bound kernels that under-utilize DRAM bandwidth may remain unaffected with less allocated bandwidth. The scaling characteristics depend both on the data size (\cref{subfig:resource-scaling-sf}) and query properties (\cref{subfig:resource-scaling-queries}).
In summary, developers are now offered with several choices on how to schedule their workloads over GPUs, but these choice are not trivial, and requires a deep understanding of both workloads characteristics and hardware properties. 

\PP{Modeling to the rescue}
To solve the above problem, currently users must rerun representative workloads with different allocations and then choose the most suitable configuration. 
This is a tedious and expensive process.
In this paper we propose
an analysis framework based on \emph{roofline models}~\cite{roofline_williams_2009}--- a widely used tool for analyzing bottlenecks.
Our analysis framework automatically estimates the resource-performance trade-offs of database systems on GPUs 
to guide users in making informed decisions about resource allocations with small errors (shown by bars in~\cref{fig:resource-scaling-time}).
Using this framework, we discover that database queries often under-utilize GPU resources. 
We uncover an opportunity to improve overall GPU utilization and workload performance by exploiting concurrent execution---our results show speedups of up to $6.43 \times$ by using our framework to select the optimal degree of concurrency. 
Finally, our modelling framework outperforms state-of-the-art (SoTA) white-box analytical model, by leveraging runtime statistics from prior executions of queries (which often occurs in practice in analytical workloads~\cite{morpheus_jyothi_2016,autotoken_sen_2020}).

\PP{Contributions}
The key contributions of this work are:
\begin{itemize}[leftmargin=*,topsep=0pt]
    \item \textbf{Comparative Analysis of GPU database systems} (\autoref{sec:char}):
    We compare performance of queries in five GPU database systems from both database and micro-architectural perspectives. 
    We find that lazy result caching, avoiding unnecessary algorithmic complexity, and eliminating unnecessary materialization of intermediate results are important for improving query performance.
    \item \textbf{Performance Modeling} (\autoref{sec:model}):
    We explore modeling query performance for two scenarios: data size change and resource change.
    For the first scenario, we significantly improve upon a SoTA model (\crystal)
    for recurring queries 
    by incorporating more realistic assumptions about memory resource utilization, which include both caches and DRAMs.
    For the second scenario, we present a roofline model that is effective in identifying and visualizing bottlenecks.
    The efficacy of the model stems from its consideration of L2 cache bandwidth along with the traditional memory bandwidth and compute resource bottlenecks.
    \item \textbf{Model-Driven Resource Management and Scheduling} (\autoref{sec:con}):
    We demonstrate that our framework is able to predict workload performance at different resource allocations and degrees of concurrency. 
    We will show how users and system administrators can automatically derive cost-performance trade-off estimates for various scheduling scenarios, eliminating the need for re-running the workload under a multitude of configurations.
\end{itemize}

%% file: background.tex
\section{GPU DATABASE SYSTEMS}~\label{sec:gpudb}
In this section, we provide an overview of the GPU database systems that
we are going to examine in this paper.
\cref{tb:gpudb} summarizes the key characteristics of these five database systems~\cite{crystal_shanbhag_2020,heavydb_2022,blazingsql_2022,surakav_he_2022,pgstrom_2022}.
These systems range from \crystal, a highly-optimized academic prototype with limited query coverage, to \heavydb, 
a production-grade system that supports a broader range of queries.
\input{tables/tb_background_system.tex}

\PP{\crystal}
\crystal~\cite{crystal_shanbhag_2020} is a recently proposed SoTA GPU database system that delivers superior query execution performance compared to other systems.
However, the types of queries that it supports are limited.
It currently only supports queries from the
Star Schema Benchmark (SSB)~\cite{ssb}.
All the queries are written in CUDA, and they have hard-coded parameters for the size of the hash table, 
the number of groups, and the size of the output table~\cite{crystalgithub_shanbhag_2020}.
While this is feasible for these queries in SSB, it is not feasible to pre-determine these parameters for arbitrary SQL queries.
\crystal also assumes that each column is a binary array generated during the pre-processing phase.
If a column contains string values, this is converted to a binary array using dictionary encoding.
This encoding process happens in CPU before query execution.

\PP{\heavydb}
\heavydb~\cite{heavydb_2022} is a widely-used GPU database system that supports many types of SQL queries.
Besides the query executor, it contains other components like query parser and query optimizer.
\heavydb takes various data formats as input, such as CSV and Parquet.
Internally, it uses LLVM~\cite{llvm_lattner_2004} to generate PTX~\cite{ptx_nvidia_2022} to build the query execution.
Additionally, \heavydb leverages a custom LLVM pass for more flexible and optimized execution strategies.
Though \heavydb currently only supports NVIDIA GPUs, it can be extended to other hardware platforms using LLVM.

\PP{\blzsql}
\blzsql~\cite{blazingsql_2022} is another open-sourced GPU database system.
Similarly to \heavydb, \blzsql is also very flexible and handles different types of queries.
Unlike \heavydb, it uses Thrust~\cite{thrust_nvidia_2022} and cuDF~\cite{rapids_nvidia_2018} as its backend for query execution.
Reusing the APIs of Thrust and cuDF has both pros and cons.
While less engineering is needed to support various operations, due to limited functionalities of existing APIs, certain operators cannot be implemented in an efficient way (\eg cannot avoid unnecessary generation of intermediate results).
\blzsql also supports different input data formats.

\PP{\pgstrom}
\pgstrom~\cite{pgstrom_2022} is an extension to the widely-used PostgresSQL database system.
It supports offloading of some operators, such as join and aggregation, to the GPU.
%
%
\pgstrom uses CUDA as its implementation backend. 
The unique characteristic of \pgstrom is that it executes queries across both CPU and GPU.
This allows the extension to be very general-purpose since it can
always fall back to CPU execution. 
On the other hand, this approach also leads to sub-optimal performance due to the overhead of data movement between CPU and GPU (as we will see shortly).
Except for \pgstrom, all the other GPU database systems that we consider are designed for GPU-only query execution where
tables are lazily cached in the GPU device memory to avoid table movement overhead during query execution.

\PP{\surakav}
\surakav~\cite{surakav_he_2022} is a recently presented GPU database system from Microsoft.
It is designed to be general purpose (\eg it supports the full TPC-H benchmark).
The interesting aspect of \surakav is that internally it uses the PyTorch~\cite{pytorch_paszke_2019} framework as its backend for executing
relational operations.
This design choice allows it to quickly support many different operations with existing GPU kernels that are already optimized.
As the PyTorch framework already supports various hardware platforms (\eg NVIDIA GPUs, AMD GPUs), \surakav inherits the portability and extensiblity of the PyTorch framework.

\PP{Query Optimization in GPU Database Systems}
All the systems have query optimization and query compilation phases except for \crystal: here
the query plans are hard-coded and pre-determined based on the selectivity of each operator (\ie selective operators are executed earlier).
\heavydb and \blzsql rely on the Apache Calcite framework for query optimization.
After query optimization, \heavydb uses LLVM to compile the query plans, and also uses LLVM to
further optimize the physical execution of GPU code. 
In contrast, \blzsql constructs the physical query plan using Thrust and cuDF libraries.

\surakav uses SparkSQL~\cite{sparksql_armbrust_2015} to optimize the SQL queries, and then translates the Spark SQL physical plans to an intermediate representation (IR).
Based on the IR, \surakav assembles a PyTorch program as a composition of pre-defined tensor programs, one for each operator in the IR.
The implementation is later optimized by the PyTorch compiler.
\pgstrom takes a different approach since it directly leverages the PostgreSQL query optimizer.
Specifically, it leverages a GPU-aware cost model that allows the GPU execution to be a part of the PostgreSQL query optimization phase.
Based on the relative cost, the optimizer determines which operators to offload to GPU.
This approach requires the optimizer to have a good estimate of the cost of each operator in the query plan.

%% file: tables/tb_background_system.tex
\begin{table}[t]
  \caption{\textbf{Qualitative comparison of GPU database systems } -- We consider the following characteristics.
  \textbf{Coverage}: whether the system is general purpose or only specific queries are supported.
  \textbf{Data Format}: formats for data input to the system.
  \textbf{Backend}: how data operators are compiled for execution.
  \textbf{Execution}: whether operators are executed on GPU or both on CPU and GPU.
  \textbf{Open Source}: if the system is publicly available.
  }
  \small

  \resizebox{\columnwidth}{!}{
  \begin{threeparttable}

    \renewcommand{\arraystretch}{1.3}
    \centering
    \begin{tabular}{@{}l|ccccc@{}}

      \toprule

      \textbf{System}
      & \textbf{\crystal}
      & \textbf{\heavydb}
      & \textbf{\blzsql}
      & \textbf{\surakav}
      & \textbf{\pgstrom} \\

      \midrule \midrule

      \multirow{2}{*}{\textbf{Coverage}}
      & \multirow{2}{*}{SSB~\cite{ssb} only}
      & \multirow{2}{*}{\makecell{General\\purpose}}
      & \multirow{2}{*}{\makecell{General\\purpose}}
      & \multirow{2}{*}{\makecell{General\\purpose}} 
      & \multirow{2}{*}{\makecell{Join\\Aggr.}} \\ \\

      \multirow{2}{*}{\textbf{\makecell{Data\\Format}}}
      & \multirow{2}{*}{\makecell{Binary\\array}}
      & \multirow{2}{*}{\makecell{CSV\\Parquet}}
      & \multirow{2}{*}{\makecell{CSV, DF\\Parquet}}
      & \multirow{2}{*}{\makecell{CSV, DF\\Parquet}} 
      & \multirow{2}{*}{CSV} \\ \\

      \multirow{2}{*}{\textbf{Backend}}
      & \multirow{2}{*}{\makecell{Hardcoded\\CUDA}}
      & \multirow{2}{*}{\makecell{LLVM\\to PTX}}
      & \multirow{2}{*}{\makecell{Thrust\\cuDF}}
      & \multirow{2}{*}{PyTorch} 
      & \multirow{2}{*}{CUDA} \\ \\

      \textbf{Execution}
      & GPU
      & GPU 
      & GPU
      & GPU
      & CPU+GPU \\
     
      \textbf{Open Source}
      & Yes
      & Yes 
      & Yes
      & No
      & Yes \\
      
      \bottomrule

    \end{tabular}

  \end{threeparttable}
  }
  \label{tb:gpudb}
  \vspace*{-3mm}
\end{table}

%% file: setup.tex
\section{EXPERIMENTAL SETUP}~\label{sec:expr}
We go over the hardware, query workloads, and profiling toolchains
used in this paper in~\autoref{sec:expr:hardware},~\autoref{sec:exp:query} and~\autoref{sec:expr:profile}, respectively.
We then describe the two execution scenarios that we consider in~\autoref{sec:expr:coldwarm}.

\subsection{Hardware}~\label{sec:expr:hardware}
We use an NVIDIA A100 GPU~\cite{a100_nvidia_2022} with Ampere GPU architecture, $40$GB of GPU memory, and $108$ Streaming Multiprocessors (SMs).
NVIDIA GPU hardware operates
at the \emph{warp} granularity. 
A warp consists of $32$ threads scheduled and executing together in a \emph{single instruction, multiple threads} (SIMT) fashion (\ie each thread executes the same instruction on different data but allowing divergence between threads).
Each SM can schedule up to 64 warps.
The A100 GPU supports newer features like multi-instance GPU~\cite{mig_nvidia_2022}.
We will provide more information on this feature in~\autoref{sec:con:back}.
%
%
The GPU is connected to the CPU via PCI-e $4$ protocol, which provides up to $32$ GB/s of bandwidth.
To achieve consistent performance,  we lock the GPU clock frequency at $1410$ MHz.

\subsection{Workloads}~\label{sec:exp:query}
Since \crystal currently only supports SSB~\cite{ssb}, 
in our experiments we decided to evaluate the SSB benchmark over the five different GPU database systems.
Even if SSB is simpler than other benchmarks like TPC-H and TPC-DS, its queries are complex enough that system characterization on this benchmark enables us to find several interesting insights. We are confident that these insights generalize to other workloads as well.

\cref{tb:query} presents a summary of the queries in the SSB benchmark.
Queries in group $1$ only have one join, and the results are aggregated into a single scalar value.
All the other queries have multiple joins (up to 4), and the final results are aggregated into a table with
multiple groups.
Based on our profiling, we found that $16$ is the largest scale factor that most database systems support without running into out-of-memory errors. 
So, we only present results up to scale factor $16$, whose tables sizes are near $7$ GB.

\input{tables/tb_back_query.tex}

\subsection{Profiling Toolchains}~\label{sec:expr:profile}
We extensively use NVIDIA Nsight System~\cite{nsightsystem_nvidia_2022}, 
Nsight Compute~\cite{nsightsystem_nvidia_2022}, and nvidia-smi~\cite{nvidiasmi_nvidia_2022} tools.
NSight System provides system-wide time breakdown, including time spent on data transfer,
memory allocation, kernel execution, etc.
However, we cannot study the kernel execution efficiency only based on those high-level
metrics, and that information is needed for query performance modeling.
Thus, we also use Nsight Compute to obtain detailed kernel execution metrics, which includes
achieved instruction per cycle, cache utilization, etc.
Nvidia-smi allows power monitoring.

\subsection{Warm \vs Cold Execution Scenarios}~\label{sec:expr:coldwarm}
We consider two execution scenarios for performance characterization: warm and cold.
With the \emph{warm scenario}, we assume that the data has been already loaded (or cached) in GPU memory, the device has been already warmed up, and the query has been already parsed, optimized, and compiled (\eg, the physical plan is available in the plan cache). 
With the \emph{cold scenario}, we assume that the data resides in CPU memory and needs to be transferred to the device, and we consider all the other overheads associated with query parsing, optimization, and compilation.

%% file: tables/tb_back_query.tex
\begin{table}[t]
  \caption{\textbf{Qualitative summary of the queries from the SSB benchmark} -- Key characteristics of queries
  in the SSB benchmark: 
  \textbf{Number of Joins}, the \textbf{Aggregation} type, whether they require \textbf{Sorting}, and query \textbf{Selectivity}.}
  \small

  \resizebox{\columnwidth}{!}{
  \begin{threeparttable}

    \renewcommand{\arraystretch}{1.2}
    \centering
    \begin{tabular}{@{}l|c @{\hspace{0.5\tabcolsep}} c @{\hspace{0.5\tabcolsep}} c @{\hspace{0.5\tabcolsep}} c @{\hspace{0.5\tabcolsep}} c @{\hspace{0.5\tabcolsep}} c @{\hspace{0.5\tabcolsep}} c @{\hspace{0.5\tabcolsep}} c @{\hspace{0.5\tabcolsep}} c @{\hspace{0.5\tabcolsep}} c @{\hspace{0.5\tabcolsep}} c @{\hspace{0.5\tabcolsep}} c @{\hspace{0.5\tabcolsep}} c@{}}
    
    \toprule
    
    \multirow{2}{*}{\textbf{Query Group}}
    & \multicolumn{3}{c}{Group 1}
    & \multicolumn{3}{c}{Group 2}
    & \multicolumn{4}{c}{Group 3}
    & \multicolumn{3}{c}{Group 4} \\
    
    \cmidrule(lr{1em}){2-4}
    \cmidrule(lr{1em}){5-7}
    \cmidrule(lr{1em}){8-11}
    \cmidrule(lr{1em}){12-14}
    
    & Q11
    & Q12
    & Q13
    & Q21
    & Q22
    & Q23
    & Q31
    & Q32
    & Q33
    & Q34
    & Q41
    & Q42
    & Q43 \\
    
    \midrule \midrule
    
    \textbf{\# Joins}
    & \multicolumn{3}{c}{$1$}
    & \multicolumn{3}{c}{$3$}
    & \multicolumn{4}{c}{$3$}
    & \multicolumn{3}{c}{$4$} \\
    
    \textbf{Aggregation}
    & \multicolumn{3}{c}{Sum}
    & \multicolumn{3}{c}{Group By}
    & \multicolumn{4}{c}{Group By}
    & \multicolumn{3}{c}{Group By} \\

    \textbf{Sorting}
    & \multicolumn{3}{c}{No}
    & \multicolumn{3}{c}{Yes}
    & \multicolumn{4}{c}{Yes}
    & \multicolumn{3}{c}{Yes} \\

    \textbf{Selectivity}
    & \rotatebox[origin=c]{70}{1.9e-2}
    & \rotatebox[origin=c]{70}{6.5e-4}
    & \rotatebox[origin=c]{70}{7.5e-5}
    & \rotatebox[origin=c]{70}{8e-3}
    & \rotatebox[origin=c]{70}{1.6e-3}
    & \rotatebox[origin=c]{70}{2e-4}
    & \rotatebox[origin=c]{70}{3.4e-2}
    & \rotatebox[origin=c]{70}{1.4e-3}
    & \rotatebox[origin=c]{70}{5.5e-5}
    & \rotatebox[origin=c]{70}{7.6e-7}
    & \rotatebox[origin=c]{70}{1.6e-2}
    & \rotatebox[origin=c]{70}{4.6e-3}
    & \rotatebox[origin=c]{70}{9.1e-5} \\
    
    \bottomrule
    
    \end{tabular}\vspace{-5pt}

  \end{threeparttable}
  }

  \label{tb:query}
\end{table}

%% file: perfanalysis.tex
\section{PERFORMANCE ANALYSIS}
~\label{sec:char}
We next characterize the performance of different database systems on the A100 GPU.
Along with the results, we also highlight the key insights obtained from these experiments.
We show insights for cold and warm query execution in~\autoref{sec:char:cold} and \autoref{sec:char:end2end}, respectively.
Afterwards, in sections~\ref{sec:char:compute},~\ref{sec:char:total},~\ref{sec:char:stall} we provide a deeper analysis of where the time is spent during the GPU execution.
Finally, in \autoref{sec:char:topkernel} we provide a brief overview of the kernel implementation efficiency.

\subsection{End-to-End Cold Query Execution}~\label{sec:char:cold}
\begin{figure}[t]
  \begin{subfigure}[t]{0.95\columnwidth}
    \centering
    \includegraphics[width=\columnwidth]{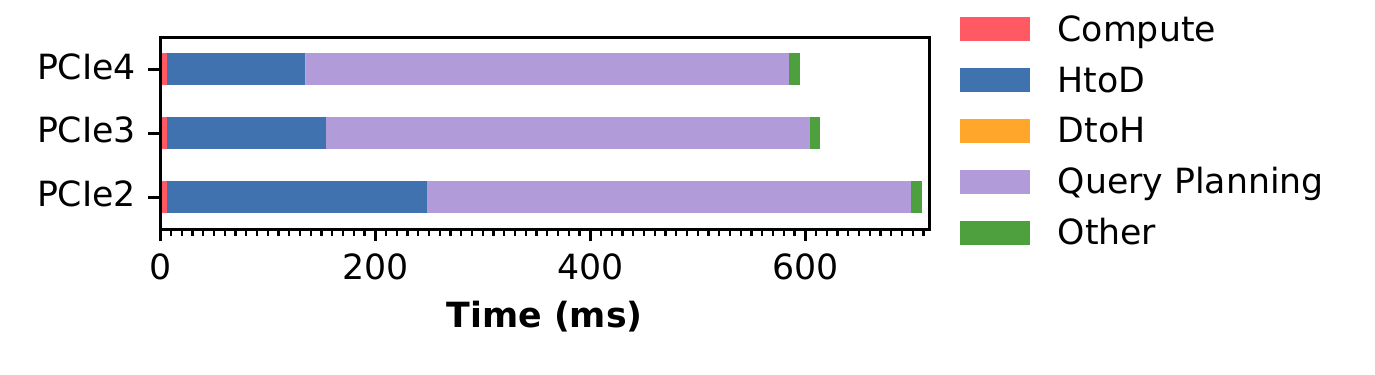}
  \end{subfigure}
  \caption{\textbf{Cold execution characterization over Q21 at SF=16 in \heavydb} -- We split the end-to-end performance
    of cold query execution into six components: \emph{Compute} time, \emph{Host to Device} data transfer (HtoD), \emph{Device to Host} data transfer (DtoH), \emph{Query Planning} time, and \emph{Other} CUDA context setup and memory management time. Similar results apply for other queries and database systems.}\vspace{-5pt}
  \label{fig:char:cold:end}
\end{figure}
With cold query execution, besides actual query execution, there are several non-negligible
overheads from both query optimization and compilation, and data transfer.
\cref{fig:char:cold:end} shows that these overheads are always greater than the overhead of the actual query execution.
As data needs to be moved to the GPU before query execution, the data transfer operation incurs a significant overhead.
During data transfer, it is not optimal to naively copy over all columns of the table.
Instead, most systems only copy the columns needed for the query to GPU, which reduces the host to device data transfer overhead.
HtoD overhead in~\cref{fig:char:cold:end} reflects the time spent on copying the required columns.

Systems also need to move results from the device back to the host.
However, as all the queries compute aggregated results that are very small in size, the device to host overhead is minimal in general.
Query plan optimization and compilation also introduces non-trivial overheads.
This is unavoidable when the system receives a query for the first time.
Most systems implement plan caching, so recurrent queries do not have the same overhead.
Lastly, we observe that the evaluated GPU database systems do not fully utilize the PCIe4 bandwidth.
While running the database systems
over PCIe3\footnotemark demonstrates significant benefit over PCIe2\footnotemark[\value{footnote}]\footnotetext{emulated by setting PCIe configuration registers.}, we found through our profiling that the benefit of moving from PCIe3 to PCIe4 
is smaller because of the under-utilization of the PCIe bandwidth.
Additionally, we found that the achieved data movement bandwidth drops when the table size does not fit into GPU DRAM capacity, because both
input data and results need to streamed in and out GPU during query execution.
\begin{highlightbox}
  \textbf{\ding{61} Finding.} \textit{Selectively transferring and caching columns to GPU, as well as 
  plan caching, are beneficial to both cold and warm query execution.} \\
  \textbf{\ding{93} Recommendation.} \textit{Improving the effective data movement bandwidth will further boost query execution performance.}
\end{highlightbox}

\subsection{End-to-End Warm Query Execution}~\label{sec:char:end2end}
\begin{figure}[t]
  \begin{subfigure}[t]{0.9\columnwidth}
    \centering
    \includegraphics[width=\columnwidth]{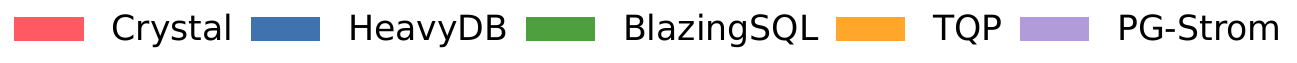}
  \end{subfigure}\vspace{-3pt}
  \begin{subfigure}[t]{\columnwidth}
    \centering
    \includegraphics[width=\columnwidth]{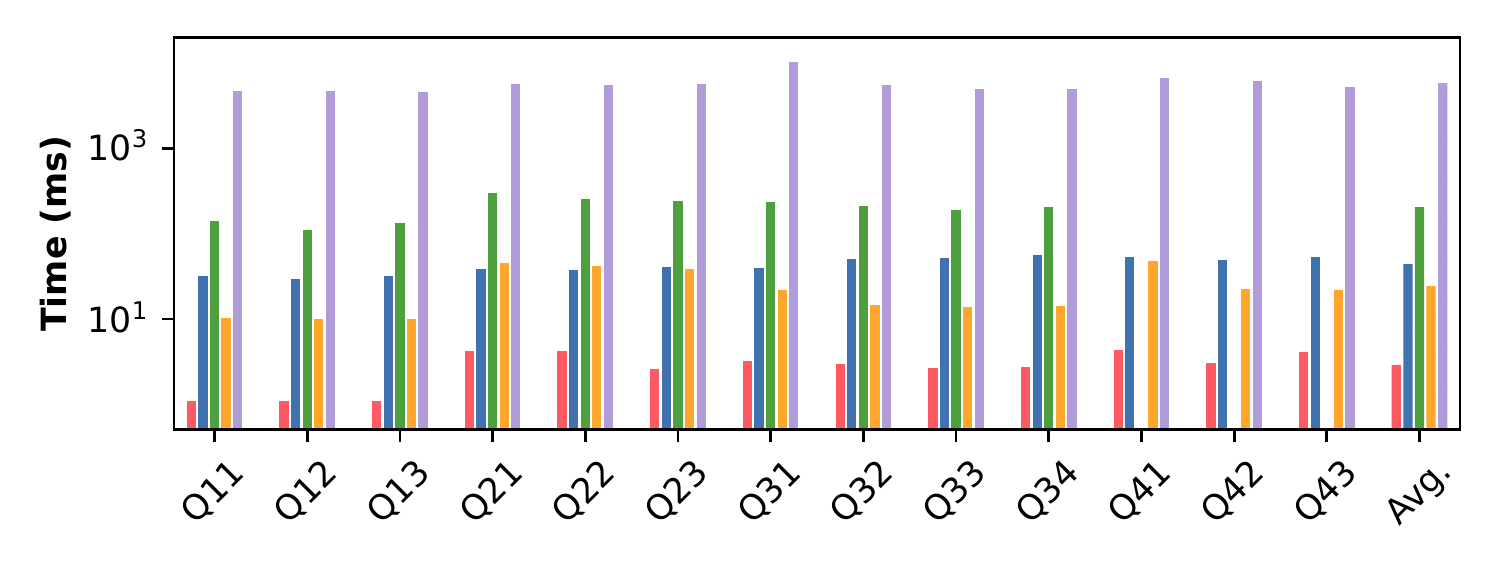}
  \end{subfigure}
  \caption{\textbf{Warm execution characterization} -- End-to-end performance for all queries (\blzsql does not support queries in group 4).}\vspace{-5pt}
  \label{fig:char:warm:end}
\end{figure}
We show the end-to-end warm query execution performance in~\cref{fig:char:warm:end}~\footnote{Most systems can successfully execute SSB queries except \blzsql, which cannot run
Q41, Q42, and Q43 due to bugs during dataframe encodings.}.
The most notable observation is that \crystal represents the upper bound in execution performance among all considered systems.
\heavydb is on average $15\times$ slower than \crystal and \surakav is $8\times$ slower
than \crystal.
\blzsql and \pgstrom are even slower.
They are $71\times$ and $2022\times$ slower than \crystal, respectively.
Next, we provide more in depth analysis on the end-to-end warm query execution of each system.

\PP{\crystal and \surakav}
For these both systems, the end-to-end time only reflects the actual GPU compute time.
In \crystal, each CUDA source file implements one query from SSB.
Once the source file is compiled, there is no additional compilation overhead associated
with the query.
Unlike \crystal, \surakav is a general-purpose system that supports a wider range of SQL queries.
In \surakav the workflow for executing queries is done in two phases: the input query 
is first parsed, optimized and compiled into a PyTorch model object; then the already-compiled model is executed over the input data.
Because to this workflow, the query execution time does not contain optimization and compilation time.
In addition, \surakav lazily caches the query results in GPU, until it is requested
by the user.
This approach is beneficial in reducing end-to-end query execution time if the results are not
immediately required, or if there is a subsequent query executing on the previously generated output results.

\PP{\heavydb and \blzsql}
For \heavydb and \blzsql, their reported end-to-end time not only consists of the GPU compute time, but also includes other overheads.
For example, \heavydb uses Calcite for query plan optimization, and LLVM to compile the query into executable code.
Since \heavydb implements a plan cache, 
we expected it to have negligible overhead for both
optimization and compilation.
However, in reality, that is not the case.
Similar to \heavydb, \blzsql also has non-trivial overheads beyond the GPU compute time.
This observation motivates us to study only the GPU compute time of the database
systems in~\autoref{sec:char:compute}.

\PP{\pgstrom}
Compared to other systems, \pgstrom has a very high end-to-end query execution time.
The reason is because of its CPU-GPU co-execution design.
In \pgstrom, operators are partially executed on CPU and partially executed on GPU.
\pgstrom needs to transfer the intermediate results from CPU to GPU or from GPU to CPU
if the execution platform changes.
Due to the design of its execution engine, 
it is challenging to implement lazy data caching
mechanism like other systems, because the intermediate results will rarely be
reused (as they are query-specific).

\begin{highlightbox}
  \textbf{\ding{61} Finding.} \ding{182} \textit{Several systems show overheads beyond the GPU query execution time.} 
  \ding{183} \textit{\pgstrom implements some operators in CPU forcing CPU+GPU co-execution.} \\
  \textbf{\ding{93} Recommendation.} \ding{182} \textit{Minimizing warm execution overheads can be major source of optimization opportunities.}
  \ding{183} \textit{CPU+GPU execution should be chosen carefully considering data transfer overhead.}
  \ding{184} \textit{Lazily copying of the query results in CPU memory can avoid unnecessary data movement overhead.}
\end{highlightbox}

\subsection{GPU Compute Time}~\label{sec:char:compute}
\begin{figure}[t]
\centering
\includegraphics[width=\columnwidth]{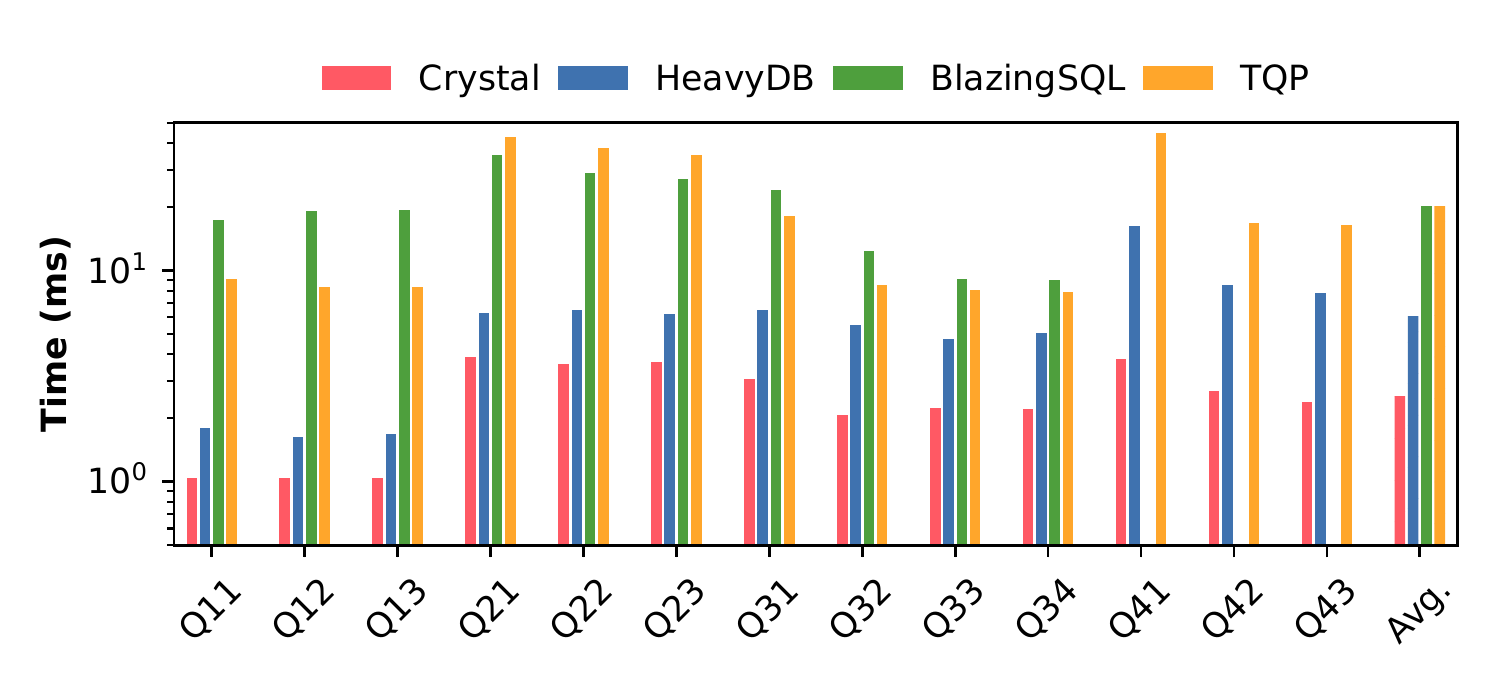}
\caption{\textbf{GPU compute time characterization} -- compared to the end-to-end warm execution of~\autoref{fig:char:warm:end}, \heavydb and \blzsql performance improve compared to \crystal, while \surakav performance decreases.}\vspace{-8pt}
\label{fig:char:gpucompute}
\end{figure}
Following the observation made in~\autoref{sec:char:end2end}, we study
the overheads of the GPU execution, for each of the considered systems.
Because \pgstrom only partially execute query plans in GPU, we do not include \pgstrom in this analysis and subsequent experiments.
The GPU compute time of the studied systems is reported in~\cref{fig:char:gpucompute}.

As expected, \crystal is still the fastest system.
However, unlike the end-to-end query time, if we only consider the GPU execution efficiency, 
\heavydb is very close to \crystal 
($2\times$ slowdown as opposed to $15\times$ slowdown reported in end-to-end comparison~\autoref{sec:char:end2end}).
\surakav is still $15\times$ slower than \crystal.
So, \heavydb is now around $4\times$ faster than \surakav.
\surakav has very similar GPU compute performance to \blzsql.
In next section, we conduct a deeper analysis of why these systems vary in their GPU execution performance.

\begin{highlightbox}
    \textbf{\ding{61} Finding.} \textit{Systems have dramatically different GPU
    compute efficiency compared to results demonstrated in end-to-end query execution time.}
\end{highlightbox}
\subsection{GPU Execution Efficiency}
~\label{sec:char:total}
We study the warm GPU execution efficiency by profiling these hardware counters:
(1) the number of integer operations, and 
(2) bytes loaded from DRAM.

\begin{figure}[t]
\begin{subfigure}[t]{\columnwidth}
  \centering
  \includegraphics[width=\columnwidth]{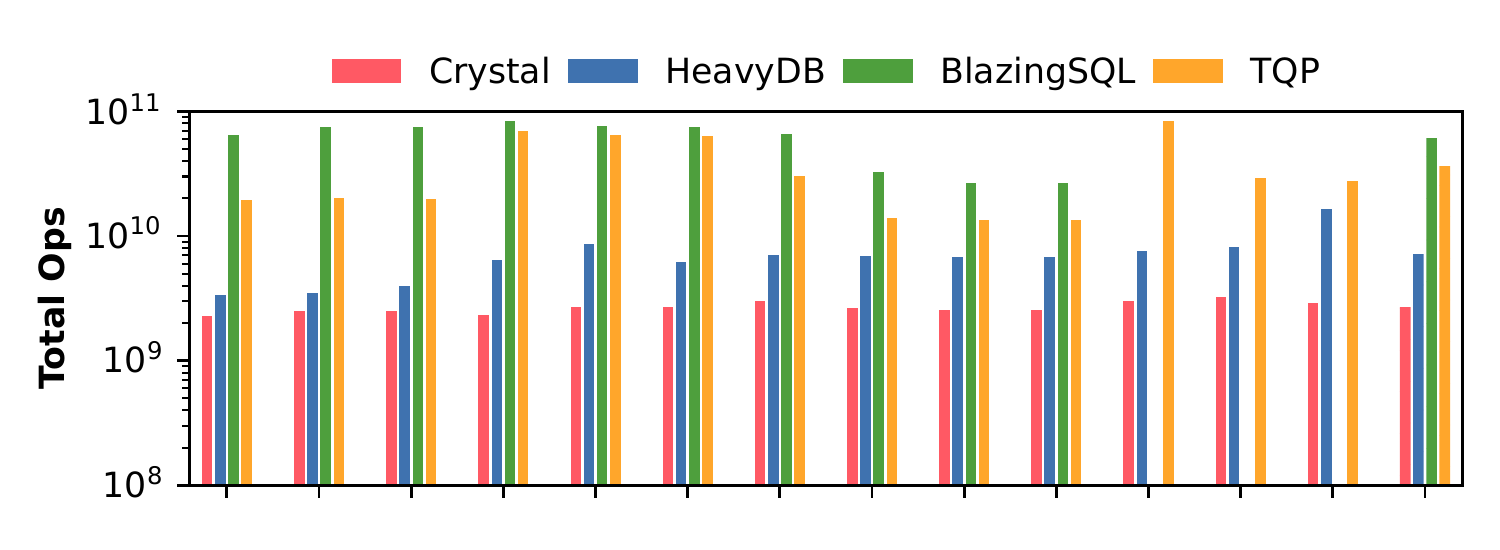}
\end{subfigure}\vspace{-5pt}
\hfill
\begin{subfigure}[t]{\columnwidth}
  \centering
  \includegraphics[width=\columnwidth]{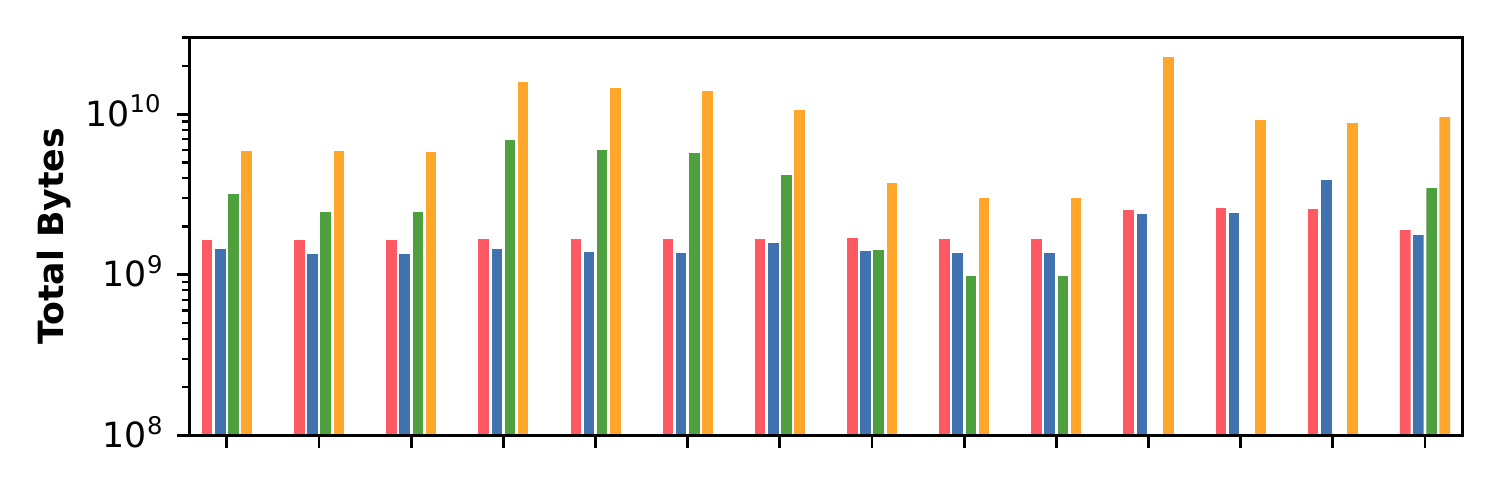}
\end{subfigure}\vspace{-5pt}
\hfill
\begin{subfigure}[t]{\columnwidth}
  \centering
  \includegraphics[width=\columnwidth]{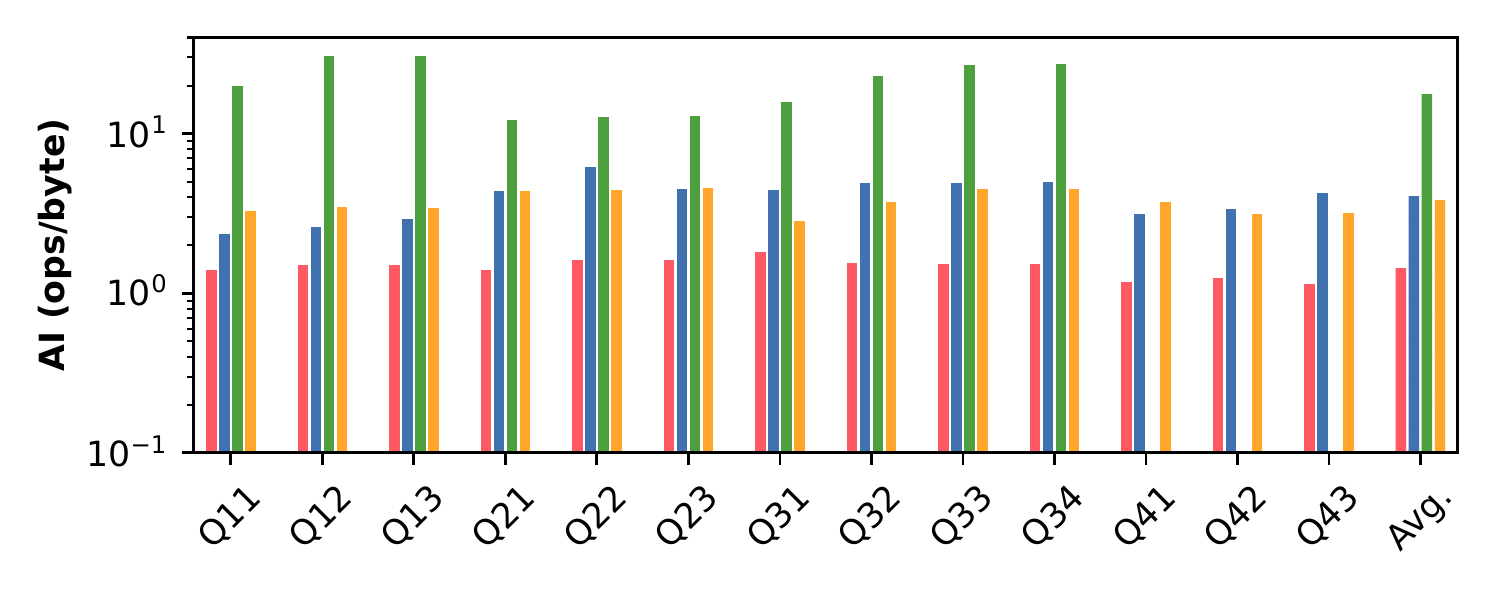}
\end{subfigure}
\caption{\textbf{GPU execution efficiency --} total integer operations (top), number of bytes from
DRAM (middle), and arithmetic intensity (bottom).}\vspace{-5pt}
\label{fig:char:detail}
\end{figure}
\PP{Integer Operations}
We first profile the total integer operations for the four systems, as
shown in~\cref{fig:char:detail} (top chart).
We find that \heavydb does slightly more integer operations than \crystal.
This is expected because \heavydb is designed to be more flexible and supports generic queries.
For example, it has GPU algorithms to calculate hash-table sizes on-the-fly, while \crystal has a hard-coded hash-table size.
On the other hand, \blzsql and \surakav uses significantly more integer operations compared to \crystal and \heavydb.
This is because those two systems have more complicated implementations, and they do not fuse operators.
For instance, \blzsql implements a very complex MurmurHash algorithm to distribute keys more evenly.

\PP{Bytes}
Next, we look into the number of bytes each system reads from the GPU device memory (\autoref{fig:char:detail}, middle chart).
Similar to number of integer operations each system operates, \crystal and \heavydb
also read less data from GPU device memory than other two systems.
We credit this to operator fusion, which allow both systems to read and materialize the minimal amount of data.
On all queries, \crystal reads slightly more data from GPU than \heavydb due to its bulk execution model.
On the other hand, \blzsql and \surakav generate many intermediate results, since they generate new columns after each query operator.
So, they read significantly more data from GPU DRAM in most of the cases.
An exception is for queries Q32, Q33, and Q34, where the amount of data loaded from GPU DRAM
by \blzsql is equal to or even lower than \crystal and \heavydb.
Our investigate reveals that \crystal does not fully avoid unnecessary data loading
after predicates.

\PP{Arithmetic Intensity}
Arithmetic intensity (AI) is a critical metric for modeling the performance of a query (number of operations divided by number of bytes read).
In~\cref{fig:char:detail} (bottom chart), we show the derived AI of the four systems.
The result suggests that \crystal, \heavydb, and \surakav have very similar AI.
In other words, all three systems perform a similar number of arithmetic operations for the same amount of data.
But, \blzsql is much more compute bound on average (around $5\times$ higher AI).

\begin{highlightbox}
  \textbf{\ding{61} Finding.} \ding{182} \textit{Number of integer operations and number of bytes read 
  are key indicators of in-GPU execution efficiency.}
  \ding{183} \textit{OLAP queries can be either compute-bound or memory-bound depending on the implementation.} \\
  \textbf{\ding{93} Recommendation.} \textit{Reducing number of bytes read by enhancing data reuse
  and avoiding unnecessary algorithmic complexity can improve GPU execution efficiency.}
\end{highlightbox}

\subsection{Warp Execution Efficiency}
\label{sec:char:stall}
The advantage GPU is that it is able to hide different types of execution stalls with its
massive parallelism.
If a warp $x$ encounters stall due to memory request, 
the GPU quickly schedules another warp $y$ for execution, so that the memory stall of $x$ is hidden.
This approach improves the number of executed instructions per cycle.
A warp may stall during execution due to various reasons.
Investigating warp execution efficiency allows us to better understand which resource bottlenecks the GPU.
Specifically, we obtain a breakdown of stalls associated with warps across systems.

\begin{figure}[t]
\begin{subfigure}[t]{0.95\columnwidth}
  \centering
  \includegraphics[width=\columnwidth]{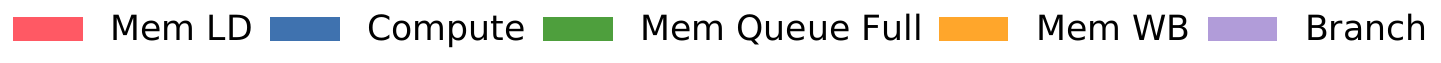}
\end{subfigure}
\begin{subfigure}[t]{0.49\columnwidth}
  \centering
  \includegraphics[width=\columnwidth]{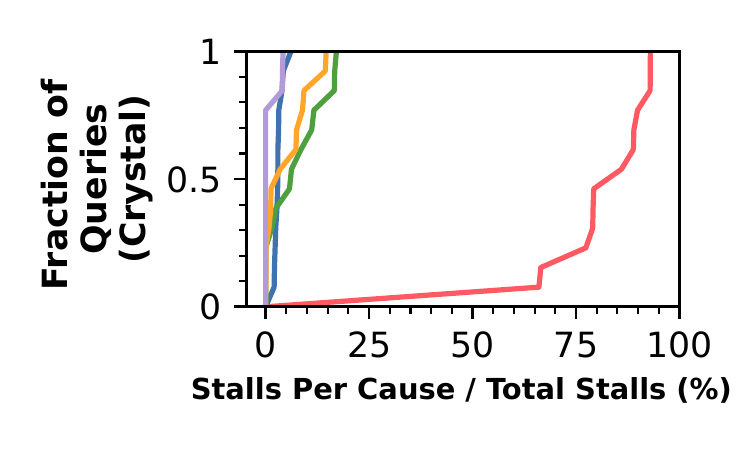}
\end{subfigure}
\hfill
\begin{subfigure}[t]{0.49\columnwidth}
  \centering
  \includegraphics[width=\columnwidth]{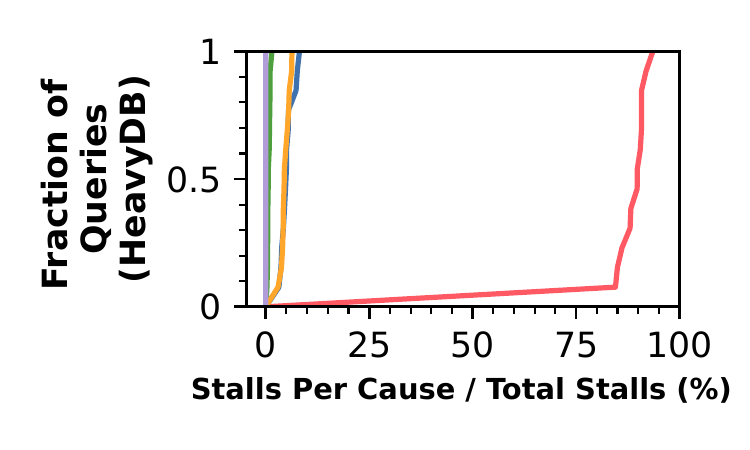}
\end{subfigure}
\hfill
\begin{subfigure}[t]{0.49\columnwidth}
  \centering
  \includegraphics[width=\columnwidth]{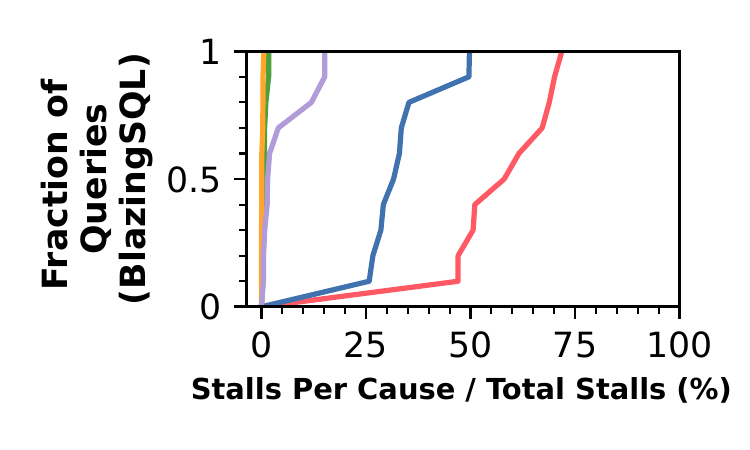}
\end{subfigure}
\hfill
\begin{subfigure}[t]{0.49\columnwidth}
  \centering
  \includegraphics[width=\columnwidth]{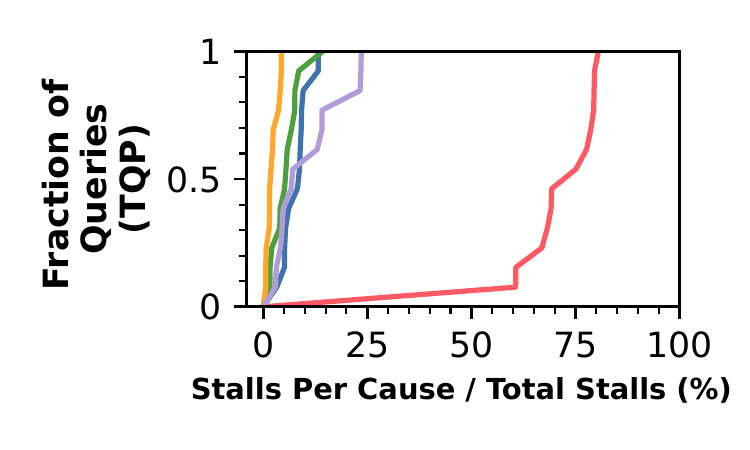}
\end{subfigure}
\hfill
\begin{subfigure}[t]{0.49\columnwidth}
  \centering
  \includegraphics[width=\columnwidth]{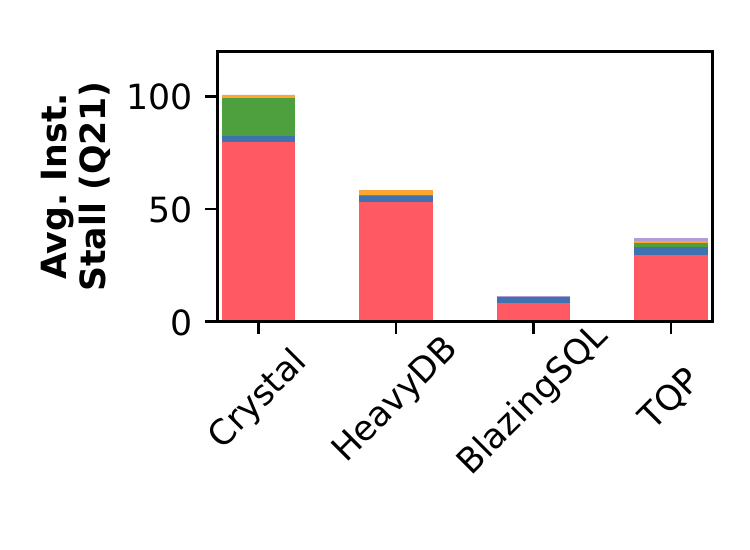}
\end{subfigure}
\hfill
\begin{subfigure}[t]{0.49\columnwidth}
  \centering
  \includegraphics[width=\columnwidth]{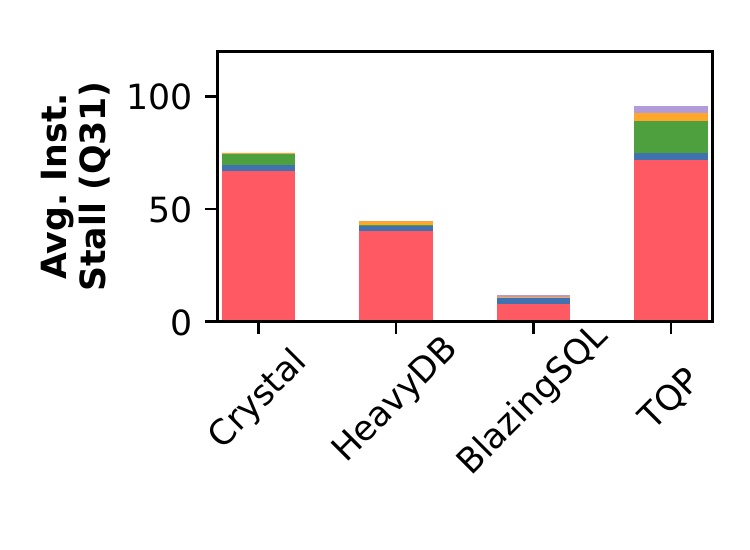}
\end{subfigure}
\caption{\textbf{Analysis of Stalls --} 
Stall distribution of all queries and detailed stall breakdown for two queries across all database systems. We only report the top five stall causes (other causes are very rare). \textsc{Mem LD}: stall for memory load request to complete. \textsc{Compute}: stall for fixed latency arithmetic operation to finish. \textsc{Mem Queue Full}: pending memory requests reach maximum. \textsc{Mem WB}: stall for results to write back to memory. \textsc{Branch}: stall due to control-divergent execution.
}\vspace{-5pt}
\label{fig:char:stall}
\end{figure}

\cref{fig:char:stall} shows the cumulative distribution functions (CDFs) of stall breakdowns for all SSB queries.
To further understand the differences between the systems, 
we pick two representative queries and show their
detailed stall breakdown (Q21 and Q31).
All other queries have stall breakdown trends similar to Q21, in which \crystal has the highest stalls, \heavydb and \surakav have lower stalls, and \blzsql always has the lowest stalls.
Stall breakdown of Q31 is the only exception, where \surakav has the highest stall compared to other due to
large intermediate results for less selective query (\autoref{sec:exp:query}).
In both stall breakdown results, \blzsql consistently has the lowest average number of warps stalled due to memory requests.
This indicates that complicated algorithms can tradeoff memory stall with arithmetic stall, but is not
beneficial for overall query performance.
\surakav tends to have more stalls when the query is less selective (\ie more rows needs to be materialized).

\begin{highlightbox}
  \textbf{\ding{61} Finding.} \ding{182} \textit{Warp stall is mostly caused by memory requests.}
  \ding{183} \textit{Using complicated algorithms ameliorates memory stalls, but does not
  improve overall performance.}  
\end{highlightbox}

\vspace*{-2mm}
\subsection{Time-Consuming Kernels}~\label{sec:char:topkernel}
\input{tables/tb_char_top_kernel.tex}
Lastly, we report a GPU compute-time breakdown for query Q21 over the four systems.
Interestingly, we found that the time breakdown between different queries is very consistent, therefore here we only show the results of query Q21.
We only report the three most time-consuming kernels due to space limitation.
We note that \blzsql and \surakav run $172$ and $190$ kernels, respectively.

\PP{\crystal and \heavydb}
These systems spend most of the query execution time in only the top kernel.
As shown in~\cref{tb:char:topkernel}, more than $90\%$ of the total compute time is spent on the top kernel.
This is because both systems implement kernel fusion, through which they execute as many operators from the query in a kernel as possible to avoid generation of intermediate results.
Because only the hash-join operation is a \emph{pipeline breaker}~\cite{pipelinebreaker_neumann_2011} for this particular query,
they implement the hash-join build phase in a separate kernel (\eg \textit{build\_htp} and
\textit{fill\_hj}).

\PP{\blzsql}
In this case, one query is split into many kernels.
Between the different kernels, generated data (\eg bitmap to indicate selected rows) needs to be materialized in order to be visible to other kernels.
This causes higher overhead compared to other systems that do not need to materialize the intermediate results.
Results show that even the hash-table probing phase already exceeds the execution time of both \crystal and \heavydb.

\PP{\surakav}
Similar to \blzsql, \surakav also executes many kernels to finish one query, and in fact it spends more time on intermediate results materialization than on actual query execution.
For example, according to the documentation from PyTorch, \textit{idx\_select} function 
(second kernel) indexes a tensor to create a new tensor.
It is used to create a new column after each predicate, but its overhead is very high compared to other kernels.
Creation of intermediate results not only causes storage overhead due to the limited GPU device memory space, but it is also costly for the entire query execution.

\begin{highlightbox}
  \textbf{\ding{61} Finding.} \ding{182} \textit{Systems with good performance require minimal number
  of kernels for a query.}
  \ding{183} \textit{Column materialization after predicate evaluation is very expensive.
  It may have even higher overhead than query operators.} \\
  \textbf{\ding{93} Recommendation.} \textit{Systems can improve performance through kernel
  fusion to avoid generation of intermediate results.}
\end{highlightbox}

%% file: tables/tb_char_top_kernel.tex
\begin{table}[t]
  \caption{\textbf{Summary of most time-consuming kernels --} three most time-consuming GPU kernels for
  executing query Q21 between all systems.}
  \small

  \resizebox{\columnwidth}{!}{
  \begin{threeparttable}

    \renewcommand{\arraystretch}{0.9}
    \centering
    \begin{tabular}{@{}llccc@{}}
    
    \toprule

    & 
    & \textbf{Top 1}
    & \textbf{Top 2}
    & \textbf{Top 3} \\
    
    \midrule \midrule
    
    \textbf{\crystal}
    & Kernel
    & \textit{probe\_ht}
    & \textit{build\_htp}
    & \textit{build\_hts} \\
    
    & Time (ms)
    & 3.96 (99.35\%)
    & 0.01 (0.37\%)
    & 0.01 (0.37\%) \\ \\
    
    \textbf{\heavydb}
    & Kernel
    & \textit{multifrag}
    & \textit{fill\_hj}
    & \textit{init\_hj} \\
    
    & Time (ms)
    & 5.86 (91.78\%)
    & 0.51 (7.96\%)
    & 0.01 (0.19\%) \\ \\
    
    \textbf{\blzsql}
    & Kernel
    & \textit{probe\_ht}
    & \textit{comp\_hj\_output}
    & \textit{parallel\_fn} \\
    
    & Time (ms)
    & 13.42 (38.26\%)
    & 12.27 (34.97\%)
    & 5.55 (15.83\%) \\ \\
    
    \textbf{\surakav}
    & Kernel
    & \textit{collect\_fn}
    & \textit{idx\_select}
    & \textit{gather\_fn} \\
    
    & Time (ms)
    & 16.88 (40.88\%)
    & 9.55 (23.13\%)
    & 5.72 (13.86\%) \\
    
    \bottomrule

    \end{tabular}

  \end{threeparttable}
  }

  \label{tb:char:topkernel}
\end{table}

%% file: perfmodel.tex
\vspace*{-3mm}
\section{PERFORMANCE MODELING}~\label{sec:model}
The performance analysis so far is based on runtime metrics observed for a given execution of the queries with a fixed input size and GPU resource allocation.
However, such analyses are unable to \textit{project} performance or analyze scalability in unseen scenarios
(\eg different input size or resource allocation).
Performance models are essential tools that enable such analyses, in addition to providing additional insights into performance bottlenecks.

In this section we present two models --- a white-box model (\autoref{sec:model:crystal}) that is dependent on the implementations of query operators and it can estimate query performance for input size changes; and a black-box model (\autoref{sec:model:roof}) that is agnostic of the query implementation details, and it can predict query performance for GPU resource allocation changes.

\subsection{White-Box Model (\crystal)}~\label{sec:model:crystal}
\crystal~\cite{crystal_shanbhag_2020} proposes a white-box model for estimating query performance on GPUs. 
It assumes that memory bandwidth is \emph{the only potential bottleneck} for OLAP query execution on GPUs, so performance are proportional to the GPU memory peak bandwidth in this model.
For example, the runtime of doing simple projection of an $int32$ column with N rows can be
estimated as follows.
\begin{align*} 
    \frac{4 \cdot \mathrm{N}}{\mathrm{Bandwidth_{\scaleto{DRAM}{4pt}}}}
\end{align*}
This model may be extended to other operators like hash join.
For example, the runtime of probing an $int32$ column with N keys in a $60$ MB hash-table on  the
A100 GPU (whose L2 cache size is $40$ MB) can be estimated as follows.
\begin{align}
    \frac{4 \cdot \mathrm{N}}{\mathrm{Bandwidth_{\scaleto{DRAM}{4pt}}}} + \frac{(1 - \frac{40}{60}) \cdot 4 \cdot \mathrm{N}}{\mathrm{Bandwidth_{\scaleto{DRAM}{4pt}}}} \label{eq:crystal:hj}
\end{align}
The left and right terms represent the time taken to load the probe table and to probe the hash-table, respectively. 
$1 - \frac{40}{60}$ indicates the potential miss rate when the kernel probes the hash-table.

\begin{figure}[t]
\begin{subfigure}[t]{\columnwidth}
  \centering
  \includegraphics[width=0.6\columnwidth]{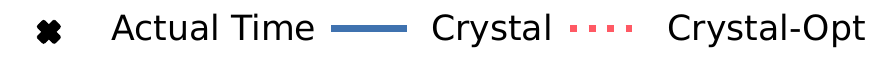}
\end{subfigure}
\hfill
\begin{subfigure}[t]{0.32\columnwidth}
  \centering
  \includegraphics[width=1.1\columnwidth]{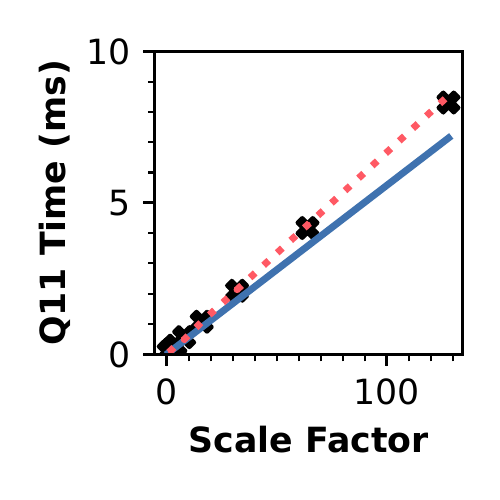}
\end{subfigure}
\hfill
\begin{subfigure}[t]{0.32\columnwidth}
  \centering
  \includegraphics[width=1.1\columnwidth]{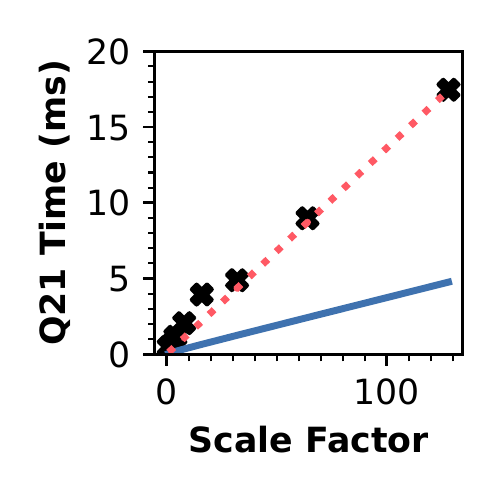}
\end{subfigure}
\hfill
\begin{subfigure}[t]{0.32\columnwidth}
  \centering
  \includegraphics[width=1.1\columnwidth]{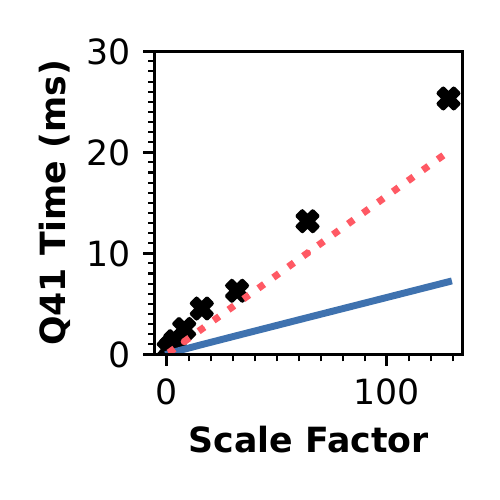}
\end{subfigure}
\caption{\textbf{Estimation Time Comparison --} 
Actual query execution time and estimated query execution time of \crystal and {\crystal}-Opt models.}
\label{fig:model:crystal-opt}
\end{figure}

\cref{fig:model:crystal-opt} shows that depending on the query and scale factor, the errors of the \crystal model can be large.
For example, Q11 in \crystal only has a scan.
So the absolute difference between estimated and actual time is small.
But, on queries with joins, such as Q21 and Q41, the gap between estimated and actual time is large.

\PP{Improving the Model for Recurring Queries}
A reason for this gap in \crystal is that it does not account for less-than-full bandwidth utilization of DRAM and cache. 
We improve the model by providing higher accuracy for warm recurrent queries by utilizing run-time statistics from previous executions.

We illustrate our improved model, {\crystal}-Opt, using two examples.
For simple projection queries, we add DRAM utilization factor (obtained from performance counter) into the equation:
\begin{align}
    \frac{4 \cdot \mathrm{N}}{\mathrm{Bandwidth_{\scaleto{DRAM}{4pt}}} \cdot \mathrm{Utilization_{\scaleto{DRAM}{4pt}}}} \label{eq:crystalopt:proj}
\end{align}
To model the runtime of hash-join probing, since the first term in~\cref{eq:crystal:hj} is for the projection, so we add the utilization factor to improve it (as~\cref{eq:crystalopt:proj}).
We then improve the second term (\ie hash-table probing) in~\cref{eq:crystal:hj} as follows:
\begin{align*}
  (1 &- \mathrm{HitRate_{\scaleto{L1 Cache}{4pt}}}) \cdot \frac{\mathrm{Cacheline_{\scaleto{L1 Cache}{4pt}}} \cdot \mathrm{N}}{\mathrm{Bandwidth_{\scaleto{L2 Cache}{4pt}}}} \\ &+ (1 - \mathrm{HitRate_{\scaleto{L2 Cache}{4pt}}}) \cdot \frac{\mathrm{Cacheline_{\scaleto{L2 Cache}{4pt}}} \cdot \mathrm{N}}{\mathrm{Bandwidth_{\scaleto{DRAM}{4pt}}}}
\end{align*}
We consider the L1 hit rate (assuming its latency is negligible) first.
We calculate the L2 access latency based on that.
Then, we calculate total data accesses to GPU DRAM considering hit rates of L1 and L2 caches.
A limitation of our model is that we need to profile the query once to obtain those performance counters (\eg L1 and L2 cache hit rates)
to predict for other scale factors.
In contrast, the original \crystal model does not require any profiling.

\cref{fig:model:crystal-opt} shows the improvement in prediction accuracy for the selected queries. 
For these predictions, we use the performance counters  from a previous run with SF=16.
For Q11, the improvement is small because \crystal implements it using scan and filter
operations, which easily saturate the DRAM bandwidth.
Taking DRAM utilization into consideration does not have significant benefits.
However, for Q21 and Q41, {\crystal}-Opt shows smaller error (max of $4.8\times$) compared to the \crystal model.
The error tends to increase for our improved model for very small or large SFs.
This is because we only profile performance counters once and project it to queries with different SFs, 
leading to less accurate estimates.
If additional profiling cost can be tolerated, the error of our improved model will further
decrease.

While white-box models give insights into how individual operators contribute to query execution time, they are specific to a given database system. 
The \crystal model only applies to \crystal and it is tedious to develop a new model for each DBMS.
The \crystal model also does not consider bottlenecks due to hardware resources other than memory bandwidth.
Now that we have shown that resources may often be under-utilized, we seek to model utilization of
GPU resources for different queries and different systems.

\begin{highlightbox}
  \textbf{\ding{61} Finding.} \textit{Memory bandwidths are under-utilized.}
\end{highlightbox}

\subsection{Black-Box Model (Roofline)}~\label{sec:model:roof}
\begin{figure}[t]
  \centering
  \includegraphics[width=0.85\columnwidth]{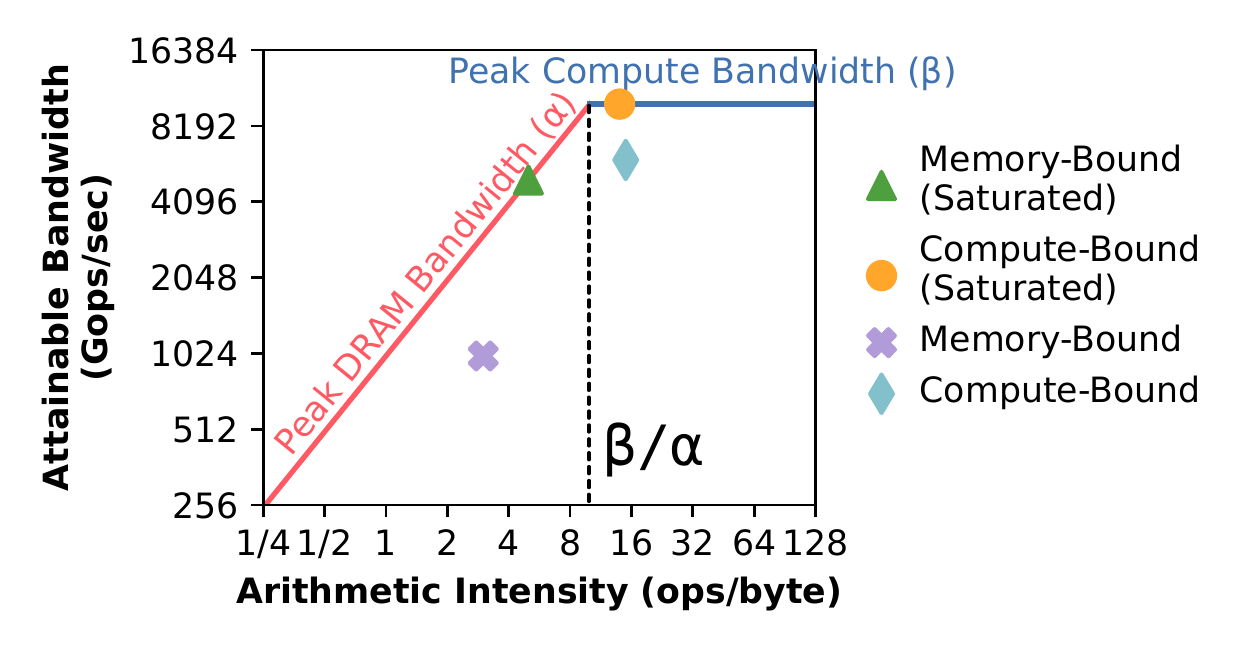}
  \caption{\textbf{Roofline Model with different
  resource bounds.}
  }\vspace{-5pt}
  \label{fig:back:roofline}
\end{figure}
The roofline model~\cite{roofline_williams_2009} assumes that any execution on a specific
hardware is bounded either by its memory resources or its compute resources.
Visually, as shown in~\cref{fig:back:roofline}, the model contains two lines to indicate the peak memory bandwidth ($\alpha$) and the peak compute bandwidth ($\beta$).
All executions on this particular hardware will correspond to points within the space bounded by these lines---those two lines are considered as the performance ceilings for that hardware.
The X-axis represents the arithmetic intensity (AI), calculated by dividing the total number of operations (\eg integer or floating-point operations) by the total number of bytes read during execution. 
The Y-axis indicates the achieved throughput, calculated as the executed operations per second.

Conventionally, a query could be either memory-bound (AI $<$ $\frac{\beta}{\alpha}$), or compute-bound (AI $>$ $\frac{\beta}{\alpha}$)
~\cite{applyroofline_ofenbeck_2014}.
The performance of algorithms that already saturate the bandwidth of either of those resources will
be impacted by changes in allocation of the corresponding resource (\eg memory-bound saturated or compute-bound saturated
in~\cref{fig:back:roofline}).
Algorithmic or compiler inefficiencies will increase the AI and make an otherwise 
memory-bound execution compute-bound, with the query taking more time to complete.
However, we show that it is important to consider \emph{L2 cache bandwidth as an additional bound} for query executions in GPUs, and that \emph{compute-bound executions are impacted by changes in compute resource allocations}. 

\PP{Case Study}
\begin{figure}[t]
\centering
\includegraphics[width=0.8\columnwidth]{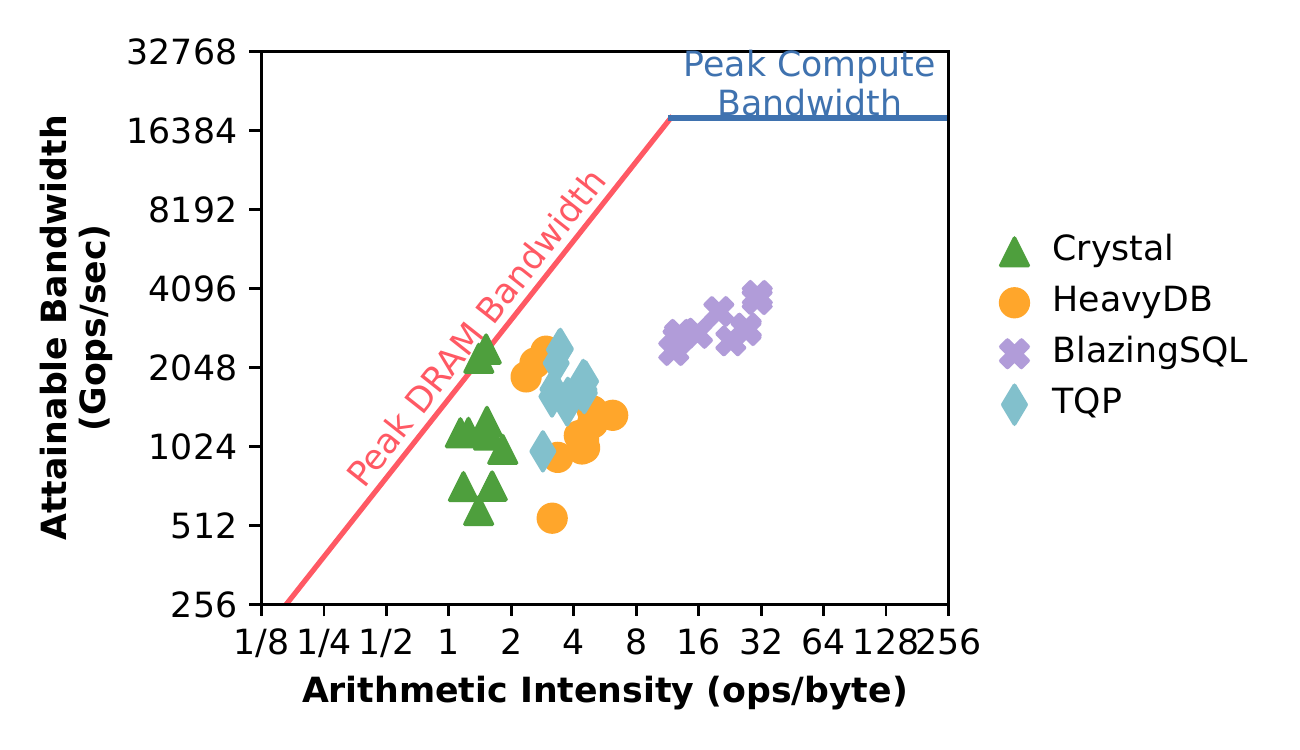}
\caption{\textbf{DRAM Roofline Model --} for SSB queries (SF=16).}\vspace{-7pt}
\label{fig:model:roof:dram}
\end{figure}
We now present our performance modeling results over the four systems in~\cref{fig:model:roof:dram}.
We obtain the metrics described 
in~\cref{tb:roof:dram}, from which we derive the AI and the attainable bandwidth.
We use the theoretical GPU DRAM bandwidth~\cite{a100_nvidia_2022} for the DRAM bandwidth ceiling.
We note that GPUs also have other functional units like floating-point operations unit.
However, because most OLAP queries only require integer operations, those functional units are not needed for constructing the roofline model.

The challenge for this type of modeling is that each query consists of multiple kernels.
Our approach is to aggregate scalar metrics such as the execution duration, total bytes, and total integer operation instructions.
We then use the aggregated metrics to obtain the required metrics for constructing the roofline model.

\input{tables/tb_roof_dram.tex}

\cref{fig:model:roof:dram} shows where all $13$ SSB queries, running on the four GPU database systems, are located with respect to the roofline model.
AI of \crystal, \heavydb, and \surakav is relatively low.
Especially for \crystal, three queries have already saturated the peak GPU DRAM bandwidth.
As discussed before (\autoref{sec:char:compute}), \crystal implements the hash-join as a filter
for Q11, Q12, and Q13 (\autoref{sec:exp:query}).
For those queries, the hash-join is just a one-to-one mapping between rows from the fact table
and the dimension table.
\crystal projects the predicate selection from dimension table to the fact table, which skips the
hash-join but simply runs a predicate selection (\ie filter) on the fact table.
As the filter operator involves running a table scan, it is feasible to saturate the GPU DRAM bandwidth (unlike hash-join).
All the other queries have a hash-join
(\autoref{sec:exp:query}), which causes many random memory accesses to GPU DRAM, so their throughput is lower than the peak DRAM bandwidth.

Compared to the other three systems,
\blzsql is instead compute-bound.
This shows that even simple OLAP queries may be quite compute-intensive
depending on the query implementation in the database system.
\blzsql has the highest AI as well as attainable bandwidth.
However, as we discussed before in~\autoref{sec:char:total}, \blzsql and \surakav load
much more data than \crystal and \heavydb, therefore they have more instructions to execute, and the  runtime of those two systems is also higher. 

\begin{highlightbox}
  \textbf{\ding{61} Finding.} \textit{Very few queries with only sequential memory accesses fully saturate the GPU DRAM bandwidth (Q11, Q12, and Q13 in \crystal).}
\end{highlightbox}

\PP{L2 Cache Bound}
We discovered that the GPU DRAM bandwidth is \textit{not} the only
resource constraint during query execution.
Especially for optimized systems like \crystal and \heavydb, they more likely saturate the peak L2 cache bandwidth ($7050$ GB/s)~\cite{jia_l2cache_a100}.
This motivates us to extend the roofline modeling methodology to L2 cache as well.
In prior work, Ilic~\cite{cacheroof_ilic_2014} has also proposed to make the roofline model to be cache-aware for CPUs.

On top of the previous study, we find that \emph{the AI of different resources is very different}.
For example, when a query has a good L2 cache hit rate, most of the memory requests will be satisfied by the L2 cache.
So, the number of bytes loaded from L2 cache will be high.
As a result, the AI relative to the L2 cache is low.
On the other hand, because there are less bytes loaded from the GPU DRAM, the AI is high with a fixed number of integer operation instructions.
This requires us to have separate roofline models to characterize the same query regarding to the different memory resources.

\input{tables/tb_roof_l2.tex}

To construct the roofline model for the L2 cache, we reuse most metrics profiled
in~\cref{tb:roof:dram} except for the total bytes read from DRAM.
To estimate the bytes read from the L2 cache, we profile the number of L2 requests (shown in~\autoref{tb:roof:l2}) that 
the kernel loads and multiply that with the cacheline size per request ($128$ bytes).

\begin{figure}[t]
\centering
\includegraphics[width=0.8\columnwidth]{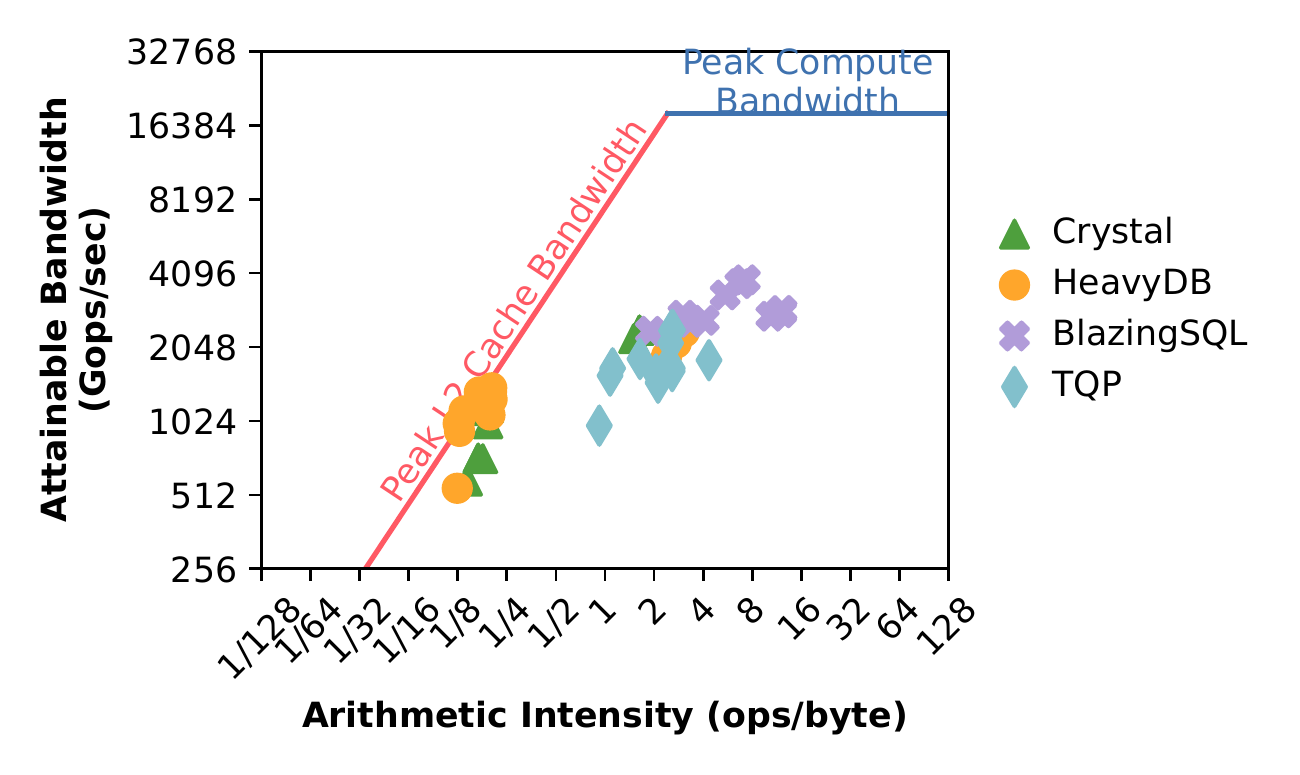}
\caption{\textbf{L2-Cache Roofline Model --} for SSB queries (SF=16).}\vspace{-5pt}
\label{fig:model:roof:l2}
\end{figure}

We present our profiling results in~\cref{fig:model:roof:l2}.
%
%
We discover that because \crystal and \heavydb are very optimized, some queries have saturated the L2 cache bandwidth.
In \blzsql and \surakav, the query implementations are still far from the L2 cache bandwidth.
In~\cref{fig:model:roof:dram}, we saw that very few queries saturate the peak DRAM bandwidth.
We observe that the AI of queries on the DRAM roofline model is lower than their corresponding AI on the L2 cache roofline model.
This is reasonable especially in the case that queries have good utilization of the L2 cache bandwidth, because most memory requests complete at L2 cache level, so the GPU has less data to handle at the DRAM level.
The second observation is that many of those queries have reached the peak L2 cache bandwidth.
Queries with hash-join are more likely to saturate the L2 cache bandwidth.
This is because SSB benchmark has relatively small hash tables that are likely to fit into the L2 cache of a high-end GPU (\eg $40$ MB).
Even though hash joins involve random accesses, queries may still exhibit good L2 cache utilization due to their small working sets.
On the other hand, queries with simple filters are more likely to be bound by DRAM bandwidth:
there is minimal data reuse, most data is streamed to the kernel,
and cache utilization is generally lower.

\begin{highlightbox}
  \textbf{\ding{61} Finding.} \ding{182} \textit{Queries with smaller working sets may be L2 cache bandwidth bound.}
  \ding{183} \textit{Good L2 cache utilization leads to lower AI for DRAM.}
\end{highlightbox}

%% file: tables/tb_roof_dram.tex
\begin{table}[t]
  \caption{\textbf{
  Profiled metrics to construct the roofline model (DRAM)}
  }
  \small

  \begin{threeparttable}

    \renewcommand{\arraystretch}{1.1}
    \centering
    \begin{tabular}{@{}ll@{}}

      \toprule

      \textbf{Metric Name}
      & \textbf{Description} \\
      
      \midrule 
      
      gpu\_\_time\_duration.sum
      & Execution duration \\
      
      dram\_\_bytes.sum
      & Total bytes from DRAM \\
      
      smsp\_\_sass\_thread\_inst\_executed\_op\_
      & \multirow{2}{*}{\makecell{Achieved compute bandwidth}} \\
      
      \hspace*{0.25em} integer\_pred\_on.sum.per\_cycle\_elapsed
      & \\
      
      sm\_\_sass\_thread\_inst\_executed\_op\_
      & \multirow{2}{*}{\makecell{Peak compute bandwidth}} \\
      
      \hspace*{0.25em} integer\_pred\_on.sum.peak\_sustained
      & \\

      \bottomrule

    \end{tabular}\vspace{-5pt}

  \end{threeparttable}

  \label{tb:roof:dram}
\end{table}

%% file: tables/tb_roof_l2.tex
\begin{table}[t]
  \caption{\textbf{Profiled metrics to construct the roofline model (L2 Cache)}
  }
  \small

  \begin{threeparttable}

    \renewcommand{\arraystretch}{1}
    \centering
    \begin{tabular}{@{}ll@{}}

      \toprule

      \textbf{Metric Name}
      & \textbf{Description} \\
      
      \midrule
      
      lts\_\_t\_requests\_srcunit\_tex\_op\_read.sum
      & Total requests to L2 cache \\

      \bottomrule

    \end{tabular}\vspace{-5pt}

  \end{threeparttable}

  \label{tb:roof:l2}
\end{table}

%% file: concurrency.tex
\section{\uppercase{Model-Driven Scheduling}}~\label{sec:con}
We have shown how to use the roofline model to identify resource
bottlenecks for different queries in~\autoref{sec:model}.
We now demonstrate how we can extend and apply the model to estimate the performance impact of changes in resource allocation and degree of concurrency.

\subsection{Resource Allocation Mechanisms}~\label{sec:con:back}
\begin{figure}[t]
\begin{subfigure}[t]{0.25\linewidth}
  \centering
  \vskip 0pt
  \includegraphics[width=0.8\columnwidth]{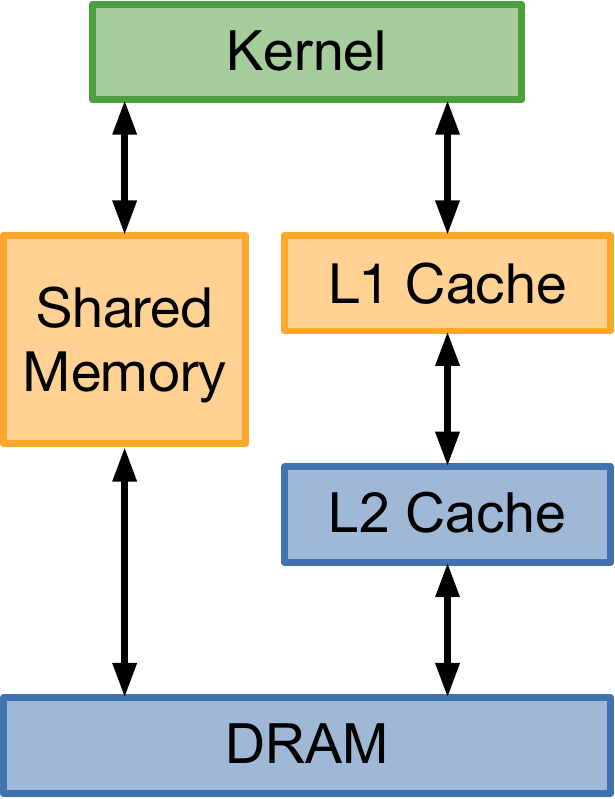}
  \caption{\textbf{GPU Memory hierarchy}}
  \label{fig:back:con:gpu}
\end{subfigure}
\hfill
\begin{subfigure}[t]{0.33\linewidth}
  \centering
  \vskip 0pt
  \includegraphics[width=0.8\columnwidth]{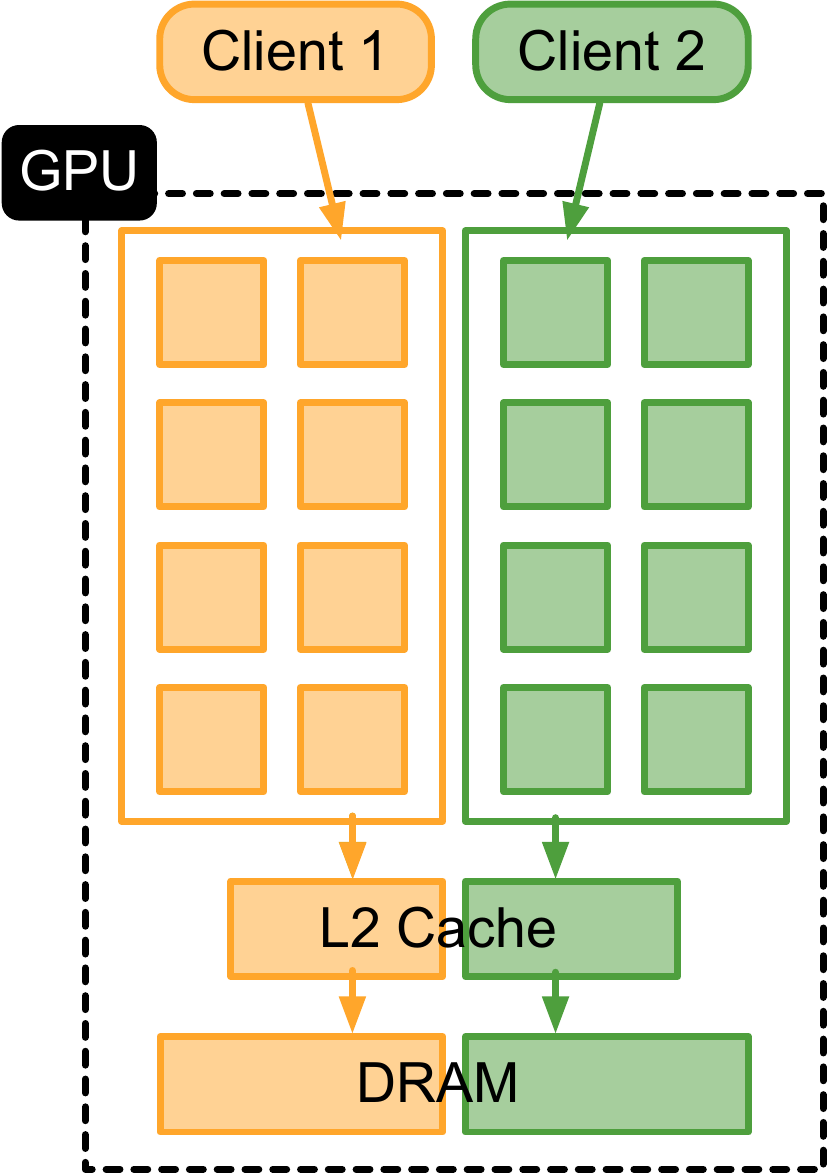}
  \caption{\textbf{MIG} -- resource partition example of multi-instance GPU (MIG).}
  \label{fig:back:con:mig}
\end{subfigure}
\hfill
\begin{subfigure}[t]{0.33\linewidth}
  \centering
  \vskip 0pt
  \includegraphics[width=0.65\columnwidth]{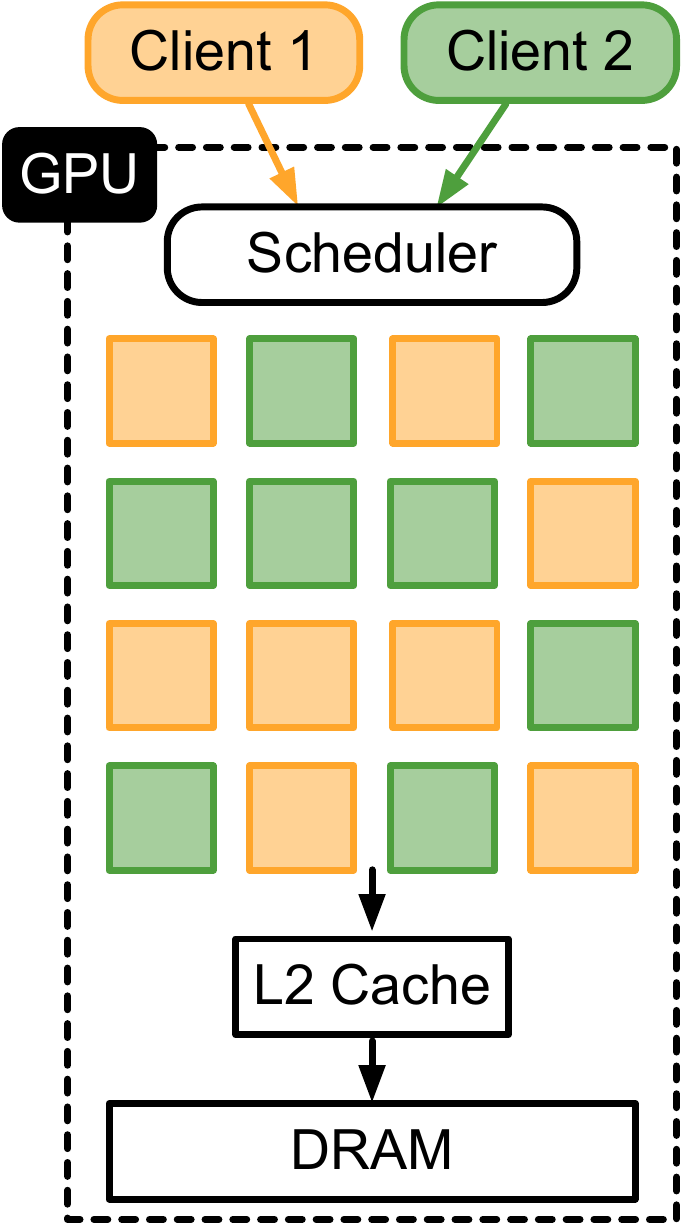}
  \caption{\textbf{MPS} -- resource partition example of multi-process service.}
  \label{fig:back:con:mps}
\end{subfigure}
\caption{\textbf{GPU Background and Concurrency Mechanisms} -- NVIDIA GPUs memory hierarchy, and 
supported concurrency mechanisms.}\vspace{-5pt}
\label{fig:gpu-concurrency}
\end{figure}
We first provide a brief overview about the memory hierarchy and currently supported
mechanisms to perform resource allocation in NVIDIA GPUs.
As~\cref{fig:back:con:gpu} shows, kernel execution can access data stored in either
shared memory or L1 cache.
Shared memory is managed completely by the user, but L1 cache is managed by the hardware.
Each GPU SM has a private L1 cache and shared memory region, 
but the L2 cache and DRAM are shared across all GPU SMs.
The memory hierarchy is consistent for all NVIDIA GPUs, but the specific values for capacity and bandwidth varies across GPUs.
NVIDIA GPUs currently support two ways for allocating resources on the GPU, and providing concurrent GPU execution capability to processes: Multi-Instance GPU 
(MIG~\cite{mig_nvidia_2022}) and Multi-Process Service (MPS~\cite{mps_nvidia_2022}).
MIG is a new feature currently supported only on the A100, A30, and H100 GPUs~\cite{mig_capability_nvidia_2022}.

\PP{MIG}
MIG enables physical partitioning of GPU resources---SMs, L2 cache, DRAM capacity and bandwidth--- which creates full isolation between concurrent processes.
\cref{fig:back:con:mig} shows an example of resource allocation through MIG to support two concurrent clients with equal allocation ($\frac{1}{2}$ GPU resources) in this example.
MIG also supports heterogeneous resource partitions to meet the different needs of
different clients.
MIG currently supports up to seven concurrent clients at its finest granularity:
each partition gets about $\frac{1}{8}$ of compute and memory resources, while nearly $\frac{1}{8}$ of the resources is reserved
for the MIG controller.
MIG currently offers a total of $18$ choices for resource partitions on the A100. 

\PP{MPS}
MPS~\cite{mps_nvidia_2022} does logical resource partitioning to support concurrent execution.
In old GPU generations, MPS only allows time-sharing of the GPU.
Since Volta, MPS allows actual concurrent execution though lightweight resource partitioning by time-sharing SMs.
In MPS, L2 cache and DRAM are still unified resources without any isolation.
When MPS starts, it creates a scheduler on the GPU.

\subsection{Model-Driven Resource Allocation }~\label{sec:con:alloc}
\begin{figure}[t]
\begin{subfigure}[t]{\columnwidth}
  \centering
  \includegraphics[width=0.4\columnwidth]{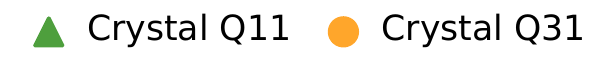}
\end{subfigure}
\hfill
\begin{subfigure}[t]{0.48\columnwidth}
  \centering
  \hspace{-2ex}
  \includegraphics[width=1.05\columnwidth]{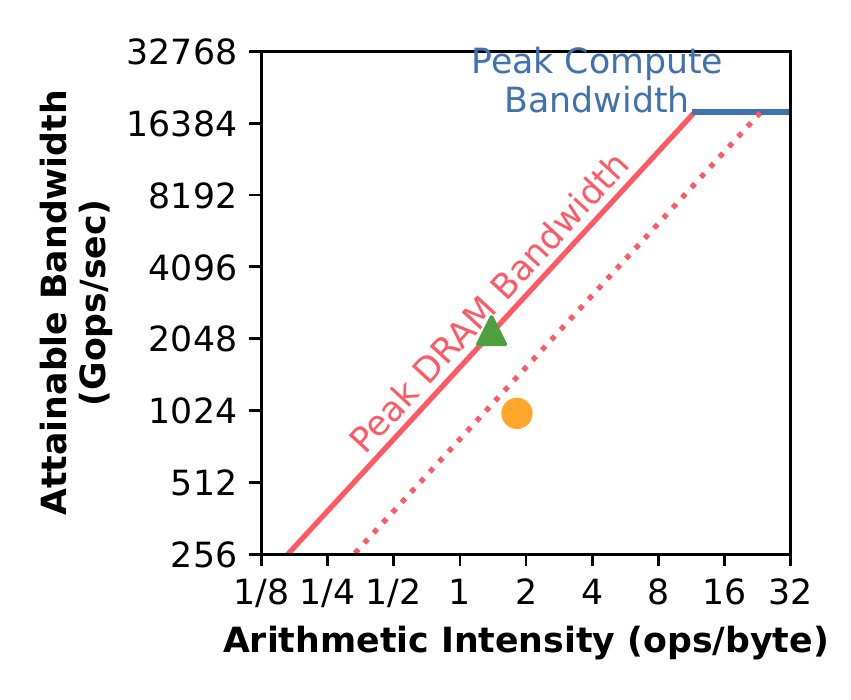}
  \caption{Query performance, memory bandwidth, and compute bandwidth of \emph{full} GPU resource 
  (\emph{dashed line is projected memory bandwidth of half GPU resource}).}
\end{subfigure}
\hfill
\begin{subfigure}[t]{0.48\columnwidth}
  \centering\hspace{-2ex}
  \includegraphics[width=1.05\columnwidth]{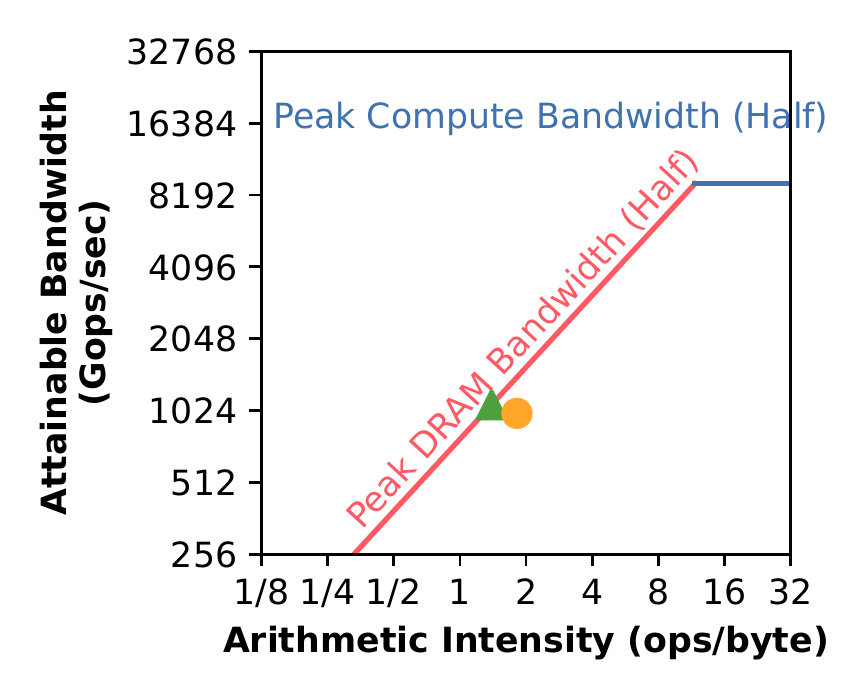}
  \caption{Query performance, memory bandwidth, and compute bandwidth of \emph{half} GPU resource.}
\end{subfigure}
\caption{\textbf{Memory-Bound Queries --} Performance impact on memory-bound queries (Q11 and Q31) for \crystal.}\vspace{-5pt}
\label{fig:con:dram}
\end{figure}
We now show how to use the roofline model to estimate performance impact for changes in resource allocation, which we can then use to select the optimal configuration. The model uses runtime statistical information which can be obtained from prior executions of recurring queries or by executing a representative workload.

We use DRAM as the target resource for illustration. 
Let $\mathrm{t}$ denote the query time under the current allocation and let $\mathrm{Bandwidth_{\scaleto{DRAM'}{4pt}}}$ denote the new DRAM bandwidth. We predict the new query time $\mathrm{t}'$ by using the following equation.
\begin{align*}
  \mathrm{t}' &= \max \left( \mathrm{t} \text{, } \frac{\text{\# of Int. Ops}}{\mathrm{AI_{\scaleto{DRAM}{4pt}}} \times \mathrm{Bandwidth_{\scaleto{DRAM'}{4pt}}}} \right)
\end{align*}
This equation is proposed based on the observation that \emph{AI is determined by the implementation of the query and is unlikely to change when the resource allocation changes.}
Our model picks the maximum among the two terms in the equation---the first term for scenarios where bandwidth is not a bottleneck (and so, the time remains unchanged), and the second for scenarios where the change in memory bandwidth hurts performance. 
For the latter case, the denominator in the fraction is the maximum throughput at the given AI and allocated bandwidth.
In this case, the query time is the time to execute the total integer operations at that throughput. 
~\cref{fig:con:dram} shows these two scenarios using representative queries (Q11, Q31) from \crystal, and full $\rightarrow$ half allocation change. Q31 under-utilizes the DRAM bandwidth and has no performance impact, whereas Q11 loses throughput (geometrically, the point shifts downwards) since it saturates the bandwidth.
Finally, we compute $\mathrm{Slowdown_{\scaleto{DRAM}{4pt}}} = {\mathrm{t}'}/{\mathrm{t}}$.

\PP{L2 Bandwidth}
As we discussed, the latest GPUs~\cite{a100_nvidia_2022} support both L2 and DRAM bandwidth partitioning.
Similar to the above method, we can estimate the impact due to changed L2 allocation using the L2 roofline model.
One challenge is how to combine both DRAM and L2 roofline models into a unified model to estimate the query slowdown.
We solve this by simply using
a max function to provide the total
slowdown estimation as follows.
\begin{align}
  \mathrm{Slowdown} = \max \left(\mathrm{Slowdown_{\scaleto{DRAM}{4pt}}}, \mathrm{Slowdown_{\scaleto{L2 Cache}{4pt}}}\right)
\end{align}
The rationale is that we empirically find that \emph{queries can be rarely bottlenecked by
both memory resources (L2 and DRAM)}.
For example, if a query has a very high utilization of the L2 cache bandwidth (\ie it is
bottlenecked by the L2), it only generates minimal traffic to DRAM,
so it will not be affected by changes in DRAM bandwidth.
Hence, one of the estimated slowdown terms is likely to be $1$ (no slowdown).
Thus, we only need to use the $\max$ function to get the dominating slowdown value.

\PP{Compute-Bound}
\begin{figure}[t]
\hfill
\begin{subfigure}[t]{0.48\columnwidth}
  \centering\hspace{-2ex}
  \includegraphics[width=1.05\columnwidth]{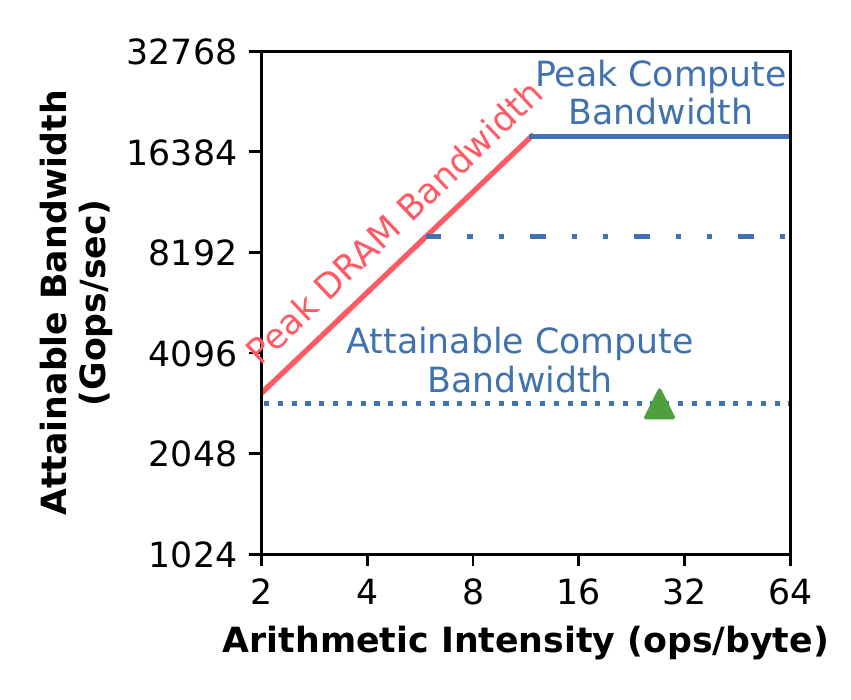}
  \caption{Query performance, memory bandwidth, and compute bandwidth of \emph{full} GPU resource 
  (\emph{dashed-dotted line projects compute bandwidth of half GPU resource}).}
\end{subfigure}
\hfill
\begin{subfigure}[t]{0.48\columnwidth}
  \centering\hspace{-2ex}
  \includegraphics[width=1.05\columnwidth]{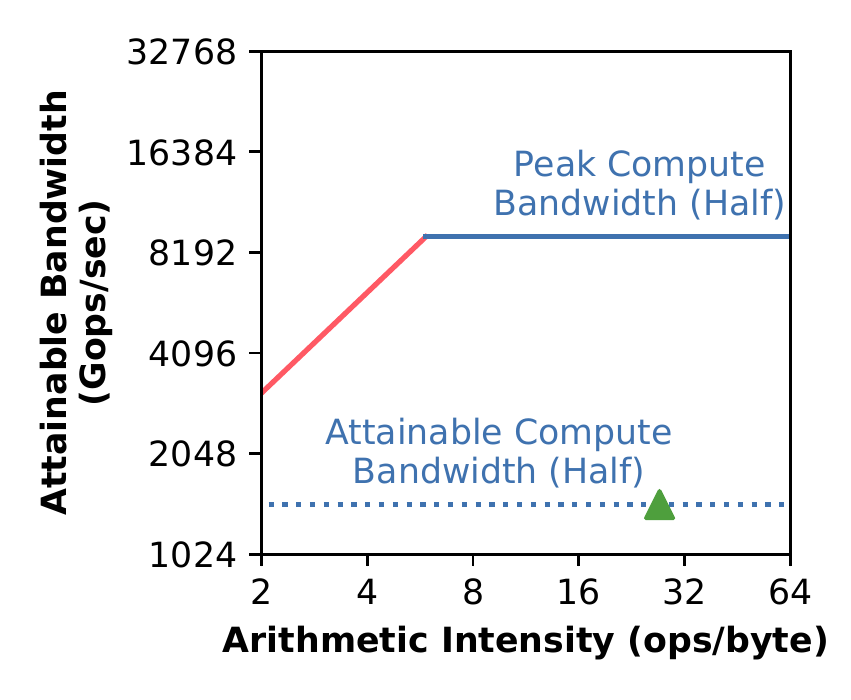}
  \caption{Query performance, memory bandwidth, and compute bandwidth of \emph{half} GPU resource.}
\end{subfigure}
\caption{\textbf{Compute-Bound Queries --} Performance impact on compute-bound queries (Q34 from \blzsql).}\vspace{-5pt}
\label{fig:con:compute}
\end{figure}
Last, we explain how we use the same model to estimate performance impact for
compute-bound queries.
In this case, because \blzsql has the most compute-bound implementations
compared to the other systems, we pick \blzsql as the example to illustrate our idea.

\cref{fig:con:compute} left shows Q34 attainable bandwidth (\ie throughput)
and AI ($2886.74$ Gops/sec and $27.15$ ops/byte respectively).
The peak compute bandwidth of the full GPU is $18247.00$ Gops/sec and that for half of GPU resources is near $9123.50$ Gops/sec (dashed and dotted line in
\cref{fig:con:compute} left).
It is clear that even the peak compute bandwidth for half GPU is beyond the attainable bandwidth of Q34, so the traditional roofline model would predict no
performance slowdown.
Nevertheless, we discover this is not the case as a result of the attainable compute bandwidth per SM being less than its peak due to execution inefficiencies (\eg memory stalls).
The overall compute bandwidth (attainable or peak) is the per-SM value $\times$ the number of SMs.
When the GPU allocates less compute resources, it reduces the number of SMs allocated, but it
cannot improve the execution efficiency (\ie attainable compute bandwidth) of each SM.
As a result, the overall attainable compute bandwidth will decrease (\cref{fig:con:compute} right).
To estimate the resulting slowdown, we can simply use the ratio of resource allocations as follows.\vspace{-5pt}
\begin{align}
  \mathrm{Slowdown_{\scaleto{Compute}{5pt}}} = \frac{1}{\mathrm{ComputeAllocationRatio}}
\end{align}
For example, if the GPU compute resources are halved, then the attainable bandwidth can
be calculated as half of the original attainable bandwidth with full GPU resources.

\PP{Unified Model}
Now that we have proposed two models for estimating slowdowns with changing allocations---one for memory resources
and one for compute resources,
the last step is to determine which model to use.
We use a simple heuristic, which is commonly used~\cite{compmembound_mao_2021,applyroofline_ofenbeck_2014}, to determine 
whether an application is compute or memory-bound.
As shown below, we can determine if an application tends to be compute-bound based on the 
AI and peak compute and DRAM bandwidths of the GPU.
The final $\mathrm{Slowdown}$
\begin{align*}
  = 
  \begin{cases}
    \text{(4)}, \text{if } \mathrm{AI_{\scaleto{DRAM}{4pt}}} > \frac{\mathrm{Bandwidth_{\scaleto{Compute}{5pt}}}}{\mathrm{Bandwidth_{\scaleto{DRAM}{4pt}}}} 
    \text{or } \mathrm{AI_{\scaleto{L2Cache}{4pt}}} > \frac{\mathrm{Bandwidth_{\scaleto{Compute}{5pt}}}}{\mathrm{Bandwidth_{\scaleto{L2Cache}{4pt}}}} \\
    \text{(3)}, \text{otherwise}  
  \end{cases}
\end{align*}
If the application is more compute-bound, then we use the model to account for compute
bandwidth reduction.
Otherwise, we apply the model for DRAM or L2 cache bandwidth reduction.

The limitation of this approach is that it can be more easily used to estimate performance impact of downsizing both memory and compute resources.
Additionally, it can be also used to reason about performance impact of upsizing compute resource.
However, it could have inaccurate estimation for performance impact of upsizing memory resources when they  are no longer a
bottleneck.

\subsection{Model-Driven Concurrent Scheduling}~\label{sec:con:degree}
Next, we extend the model to estimate end-to-end performance impact for different degrees of concurrency. This can be used to determine the optimal concurrency
for the best performance.

\PP{CPU and Constant Overhead}
To construct the model, 
we first need to consider additional overheads for query executions.
For CPU overheads, our model includes the overhead of query optimization and compilation.
For some systems (\eg \heavydb, \blzsql), even though the same query has already been
optimized and compiled to a binary, each query invocation still introduces some constant
overhead on the CPU side.
For all systems, we also consider those overheads.
We will later show insights about how concurrency is also beneficial to alleviate those overheads.
There are two additional major overheads that we consider---
GPU setup overhead, which includes GPU context initialization and memory allocations, and
data transfer overhead.
All systems cache tables on the GPU device for future query executions, so the data
transfer overhead is also only a one time cost.

\PP{End-to-End Performance}
We now can estimate the end-to-end query execution time individually for each process.
When the system varies the resources, the model will adjust the query
GPU execution time.
Other overheads will remain unchanged.
Now to consider time from multiple concurrent processes $P$, we use a $\max$ function because
the longest running process will determine the end-to-end query execution time for
concurrent executions.
\[ \mathrm{ExecTime} =  \max \left( \mathrm{ExecTime_{\scaleto{P_{1}}{4pt}}}, \mathrm{ExecTime_{\scaleto{P_{2}}{4pt}}} \text{ ... } \mathrm{ExecTime_{\scaleto{P_{n}}{4pt}}} \right) \]
Our model can also be used to estimate query performance \vs degree of concurrency for MPS (\autoref{sec:con:back}) excluding accounting for interference in accesses to the shared L2 cache and DRAM.

%% file: evaluation.tex
\subsection{Evaluation}~\label{sec:eval}
We focus on answering the following questions:
\begin{itemize}[leftmargin=*]
  \item \textbf{RQ1 -- How accurate is our model for estimating GPU query execution performance \vs resource allocation compared to a naive linear scaling model?}
  \item \textbf{RQ2 -- How beneficial is concurrent query execution?}
  \item \textbf{RQ3 -- How good is our model to estimate end-to-end query execution performance \vs degree of concurrency?} 
\end{itemize}

We demonstrate our approach using \heavydb for illustration, but expect similar conclusions to hold for other systems as well.

\PP {RQ1 -- Model Accuracy on Resource Allocation}
\begin{figure}[t]
\begin{subfigure}[t]{\columnwidth}
  \centering
  \includegraphics[width=0.7\columnwidth]{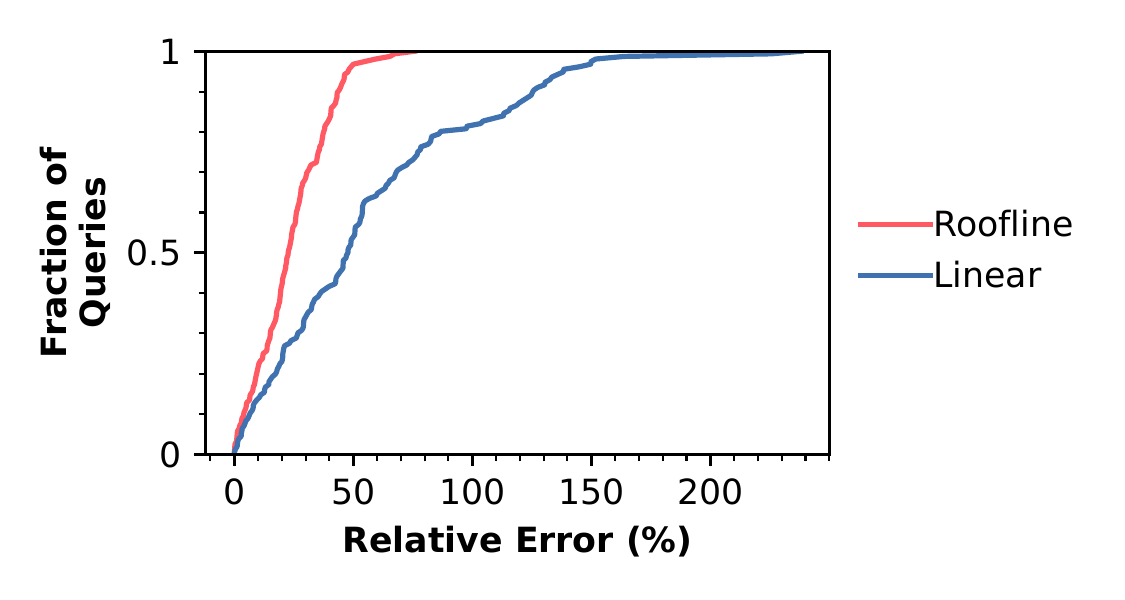}
\end{subfigure}
\caption{\textbf{Query Time Estimation Error --} Comparison of CDF of query
time estimation error between roofline and linear model.}\vspace{-5pt}
\label{fig:eval:perfrsrc-model-acc}
\end{figure}
\cref{fig:eval:perfrsrc-model-acc} shows the distribution (CDF) of relative error, which is computed as $\left(\frac{|\text{Estimate Time} - \text{Actual Time}|}{\text{Actual Time}}\right) \times 100$, for all SSB queries with four SFs
($2, 4, 8, 16$) running on \heavydb, using estimates from our roofline model and from a linear scaling model {($\mathrm{t}' = \frac{\mathrm{t}}{\mathrm{ComputeAllocationRatio}}$)} that we use as a baseline.
Both models use profiled information from a single run of each query with a given SF and with full GPU allocation to estimate query runtimes at other allocations.
We observe that \emph{our roofline model is substantially more accurate than the baseline}---the median and 95$^{th}$ percentile errors are $22.73$\% and $47.79$\% respectively in contrast to $48.39$\% and $139.77$\% for the baseline.
Queries run with large SFs or having mainly sequential scan operations are more likely to have
linear scaling with resource allocation changes.
However, our roofline model
provides better estimation accuracy for queries whose performance do not scale linearly with GPU resources (\autoref{sec:intro}).

\PP {RQ2 -- Benefit of Concurrent Query Execution}
We implemented a concurrent query execution system and designed the following experiment to evaluate benefits 
with Degree of Concurrency (DoC) = $2$, $3$, and $7$, over DoC=$1$ (no concurrency), for SSB queries on \heavydb. 
We configured the GPU, through MIG, to support the required DoC, and started the 
corresponding number of \heavydb instances, one on each GPU partition. 
We construct a simple scheduler to dispatch SSB queries in a randomized sequence 
to each \heavydb instance, repeated $1000$ times.
We then measure the overall throughput (queries per second \ie QPS) for the GPU.

\begin{figure}[t]
\begin{subfigure}[t]{\columnwidth}
  \centering
  \includegraphics[width=0.5\columnwidth]{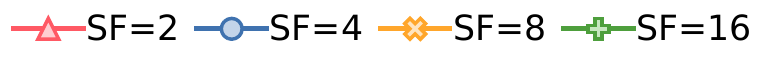}
\end{subfigure}
\hfill
\begin{subfigure}[t]{0.49\columnwidth}
  \centering
  \includegraphics[width=0.97\columnwidth]{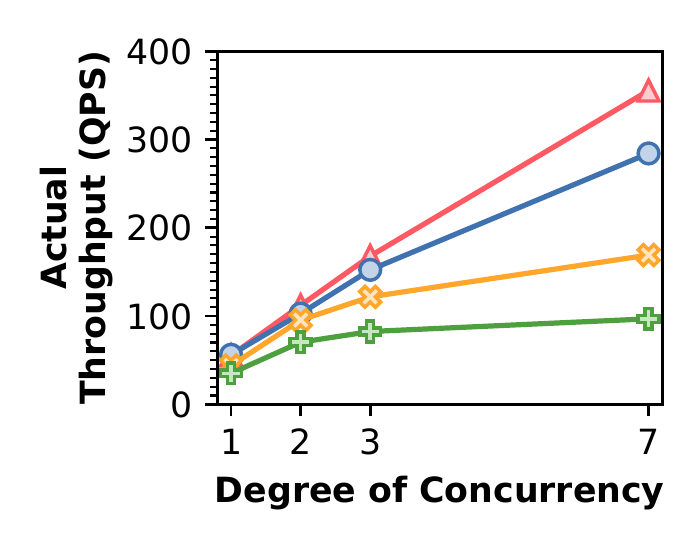}
\end{subfigure}
\hfill
\begin{subfigure}[t]{0.49\columnwidth}
  \centering
  \includegraphics[width=0.97\columnwidth]{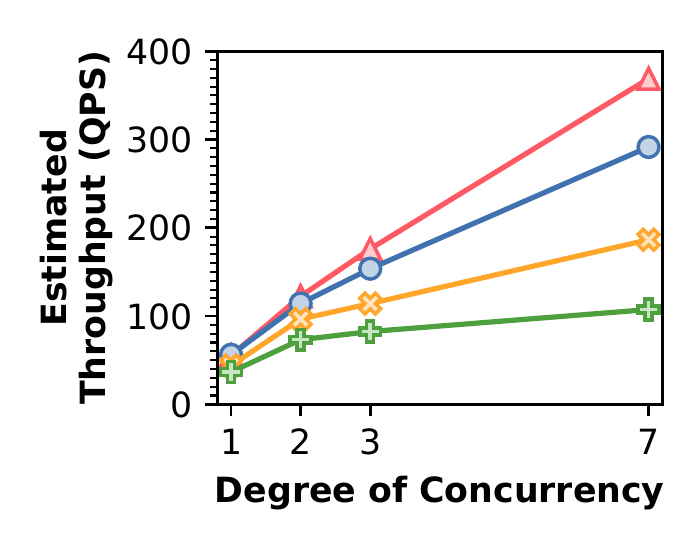}
\end{subfigure}
\caption{\textbf{Throughput \vs Degree of Concurrency --} Comparison between actual and
estimated throughput of running queries concurrently.}\vspace{-3pt}
\label{fig:eval:perfdegreecon-model-acc}
\end{figure}
\cref{fig:eval:perfdegreecon-model-acc} (left) shows the actual (measured) throughput for different DoC. We find that concurrent execution provides near-linear performance improvement---$6.43\times$ and $5.06\times$ for SF=$2$ and SF=$4$ respectively at DoC=$7$ (over DoC=$1$). Queries at small SFs typically do not fully saturate any resource. More queries can finish within a fixed time with concurrent execution, so the overall throughput increases. 
In contrast, queries are more easily affected by resource reduction at larger SFs, leading to a significant increase in query latency that overshadows the gains of concurrent execution. However, concurrency is still able to hide CPU latency overhead (\autoref{sec:char:end2end}), leading to substantial throughput improvements---$3.85\times$ and $2.75\times$ for SF=$8$ and SF=$16$ respectively at DoC=$7$ (over DoC=$1$).

\PP{RQ3 -- Model Accuracy on Degree of Concurrency}
\cref{fig:eval:perfdegreecon-model-acc} (right) shows the estimated overall throughput vs DoC for the above experimental setting. We use our models to estimate end-to-end query execution time for all DoC using profiling data from a single run of the query with full GPU allocation (DoC=$1$). Our estimated throughputs are very close to actuals, both for absolute values as well as for scaling trends (linear and non-linear).

%% file: related.tex
\vspace*{-2mm}
\section{RELATED WORK}
\vspace*{-2mm}
\PP{GPU database systems}
Suh~\etal~\cite{gpuchar_suh_2022} is the most recent study to compare different GPU database systems.
It only focuses on end-to-end execution time comparison of different
systems on different GPUs.
However, our paper shares more insights on internals of GPU execution efficiency.
Yin and Yang~\cite{yinyang_yuan_2013} studies GPU execution of SSB~\cite{ssb_rabl_2013}, which focuses
on presenting time breakdown between kernels and data transfers, but it never examines
the effciency of each kernel.
Pump up the Volume~\cite{pump_lutz_2020} and triton join~\cite{triton_lutz_2022} study performance impact 
from newer CPU-GPU interconnects on out-of-GPU hash-join.
GPL~\cite{gpl_paul_2016} also observes that the resource of GPU can be underutilized during query
execution.
Furst~\etal~\cite{gpuchar_furst_2017} studies GPU kernel efficiency but only focuses on showing GPU occupancy
\vs types of instruction comparison.
Funke~\etal~\cite{pipeline_funke_2018,dataparallel_funke_2020} and Paul~\etal~\cite{jitgpudb_paul_2020} optimize JIT
compilers for better GPU execution efficiency through kernel fusions and thread divergence elimination.
Sioulas~\etal~\cite{parthj_sioulas_2019} implements partitioned hash-join in GPU.
Shanbhag~\etal~\cite{tilecompress_bobbi_2022} also extends \crystal library for in-GPU compression.
Instead, HippogriffDB~\cite{hippogriffdb_li_2016} accelerate query performance by supporting GPU execution
directly on compressed data.
Mordred~\cite{cpugpudb_bobbi_2022} and HetExchange~\cite{hetexchange_chrysogelos_2019} both explore CPU-GPU
query executions.
Rosenfeld~\etal~\cite{gpusurvey_rosenfeld_2022} provides an in-depth overview of CPU-GPU database systems.
Gacco~\cite{gacco_boeschen_2022} instead studies transactional query processing on GPUs.
MGJoin~\cite{mgjoin_paul_2021} and Maltenberger~\etal~\cite{multigpu_maltenberger_2022} evaluates join and sorting
algorithms for multi-GPU systems.
Doraiswamy and Freire~\cite{gpuspatial_doraiswamy_2020} propose to use GPUs to process spatial data 
(\eg geometric objects) in database systems.
We cover existing GPU database systems in~\autoref{sec:gpudb}.
Our work provides in-depth micro-architectural analysis of existing systems.

\PP{GPU performance modeling}
Hong and Kim~\cite{gpuperf_hong_2009} model performance of early generation GPUs, where GPU caches
are still not mature, so the focus is on GPU DRAM and compute bandwidths.
Zhang~\etal~\cite{quantperfgpu_zhang_2011} provides performance optimization suggestions to
developers through GPU performance modeling with micro-benchmarking profiling.
Wu~\etal~\cite{mlgpuperf_wu_2015} instead proposes to use machine learning approach to predict
kernel performance, which also requires profiling performance counters on real hardware.
Baghsorkhi~\etal~\cite{codegpuperf_baghsorkhi_2010} uses code analysis to consider performance impact
from control flow divergence and memory bank conflicts.
Our approach of using the roofline models provides a simple way to analyze performance impact
for different GPU resources without requiring code analysis or machine learning models.

Gables~\cite{gables_hill_2019} uses roofline models to study SoC platforms with multiple
accelerators.
Ding~\etal~\cite{gpuroofline_ding_2019} and Lopes~\etal~\cite{gpuroofline_Lopes_2017} also apply roofline models to GPUs.
Ding~\etal captures all instruction for AI, which can lead to inaccurate estimations.
Lopes~\etal also explores the cache-aware aspect, but does not differentiate between AI at cache-level
and at DRAM-level, in contrast to our work.
Additionally, we extend and apply the roofline model to estimate query performance for different resource allocations on the GPU, that has not been explored in prior work.

\PP{Concurrent execution in GPUs}
Yu~\etal~\cite{migsurvey_yu_2022} has also surveyed the tradeoff between MIG and MPS.
The paper shares the potential opportunities and use cases of using concurrent execution in
GPUs.
Tan~\etal~\cite{dnnmig_tan_2021} explores accelerating deep neural network (DNN) inference by using
MIG.
Kass~\etal~\cite{migdl_kaas_2022} instead investigates DNN training in MIG.
MISO~\cite{miso_li_2022} uses MPS to find the best GPU partition configuration and run
the actual execution in MIG to achieve the best performance.
In our work, we support relational query operations in the form of concurrent execution, which
is often limited by the GPU memory resource.
Instead, DNN inference is more compute-bound.
Our work uses simple models to estimate performance with different GPU resource limits, whereas others~\cite{dnnmig_tan_2021} require profiling the execution for different configurations.

%% file: discuss.tex
\vspace*{-2mm}
\section{RESEARCH DIRECTIONS}~\label{sec:diss}
We conclude with a discussion about directions for research in hardware and software improvements for GPU database systems.

\PP{GPU L2 Cache Improvements}
The GPU L2 cache can commonly be a bottlenecked
resource (\autoref{sec:model:roof}).
Current GPUs adopt 
a shared L2 cache design, in which data in
GPU DRAM is mapped to different L2 cache slices based on their address in memory.
GPU SMs that access near-memory addresses frequently (\eg for probing a relatively small
hash-table) need to serialize memory accesses in the L2 cache, because the requested
cachelines are all located in the same cache slice.
Previous works~\cite{memoryside_zhao_2019} have shown that new GPU architectures with 
a private L2 cache for a cluster of SMs can provide better bandwidth.
\eat{Even if different SMs request the same cache line, they will access different cache slices
because L2 cache maintains multiple copies of the same cache line in the private L2 cache design.
Knowledge of hardware details can enable software optimizations to make query executions more GPU hardware friendly.}
Database developers can store frequently-accessed data in different 
memory partitions (consequently, mapped to different L2 cache slices) to maximize L2 cache utilization.

\PP{Efficient and Automatic Kernel Fusion}
We observed that kernel fusion is one of the most important techniques for achieving good performance by avoiding unnecessary data materialization (\autoref{sec:char:topkernel}).
Nevertheless, only \heavydb supports automatic kernel fusion, but the compilation overhead is not negligible (\autoref{fig:char:cold:end}).
\surakav leverages PyTorch for operator fusion, but the set of kernels that can be automatically fused is minimal, and mostly ML-related (e.g., convolutions with activation functions)~\cite{tvm,taso}.
On the CPU side, in the last few years several techniques have been proposed for efficient operator fusion~\cite{10.14778/3476311.3476410}. 
Applying the same techniques in the GPU space is an interesting research direction.

\PP{Flexible Resource Allocation}
MIG currently does not allow separately selecting compute and memory resource allocations, and only a set of predefined resource partitions are available to choose from.
However, results in~\autoref{sec:model:roof} show that queries have diverse resource requirements.
Decoupled partitioning of GPU resources can create more optimization opportunities for database
workloads.

\PP{GPU Resource-Aware QO}
As we discussed earlier (\autoref{sec:intro}), there is a wide variety of GPUs available today having many different sizes for various resources. Additionally, MIG capability has increased the set of possible resource allocations to run a query. Having a resource-aware query optimizer, for GPU databases, that adapts query plans to allocated resources may improve query performance compared to that with a fixed, resource-oblivious query plan.

\PP{\ding{182} Example: Memory Coalescing}
We observed that most queries have long memory stalls (\autoref{sec:char:stall}).
One way to reduce memory stalls is to increase memory coalescing~\cite{cudamemorycoalescing_nvidia_2013}.
For example, WarpDrive~\cite{warpdrive_junger_2018} demonstrates that achieving better
memory coalescing for accesses to GPU DRAM can improve performance even though it
incurs more arithmetic operations.
Nevertheless, for newer GPUs with larger L2 cache size (less than $10$ MB for previous generation GPUs~\cite{p100gpu_techpowerup_2022,v100gpu_techpowerup_2022}), 
this optimization may no longer be critical, if a hash-table fits into the L2 cache.
Instead, additional arithmetic operations caused by memory coalescing can hurt performance.

\PP{\ding{183} Example: Exploiting L1 Cache}
GPUs offer developers capability to control data caching behavior (\eg bypass L1 cache or not and 
cache replacement policy) through PTX~\cite{ptx_nvidia_2022}.
Previously, the per-SM L1 cache in old generation GPUs had longer latency and lower capacity (\eg
$24$ KB in P100~\cite{p100gpu_techpowerup_2022} in contrast to $192$ KB in A100), 
so it was skipped as optimization to reduce memory access latency
for workloads with many random accesses (\eg hash-join).
However, using the L1 cache in newer GPUs can be very beneficial, especially for skewed workloads,
{because data reuse is higher}.

\PP{Supporting Larger Datasets} While GPU memory capacities have increased, the largest size at the time of this writing is possibly 128 GB~\cite{mi250_amd_2021} which is several times smaller than CPU memory in high-end servers. Our experiments also show diminishing reductions in data movement overhead for higher PCIe generations (\autoref{sec:char:cold}). Considering these factors, it would be useful to explore how GPU database systems could efficiently process larger datasets, e.g., by using techniques such as query processing on compressed data~\cite{hippogriffdb_li_2016}.

\PP{Time Prediction for different GPUs} We have demonstrated (\autoref{sec:eval}) that our models produce good estimates of query performance for different resource limits on the same GPU architecture. Future research can explore extending this capability to predict performance for different GPU architectures as well.

%% file: conclusion.tex
\vspace*{-2mm}
\section{CONCLUSIONS}
In this work, we provide micro-architectural insights studied on existing GPU database systems.
We demonstrate two models to do query performance analysis.
Our proposed approach shows good accuracy at estimating query performance with respect to
different GPU resources and concurrency.